\documentclass{article}
\usepackage{graphicx} 
\usepackage{setspace}
\usepackage[a4paper, total={6in, 8in}]{geometry}

\usepackage{amsmath}
\usepackage{bbm}
\usepackage{amssymb}

\usepackage{caption}

\usepackage{natbib}
\usepackage{subcaption}
\RequirePackage[colorlinks,citecolor=blue,linkcolor=blue,urlcolor=blue,pagebackref]{hyperref}

\usepackage{booktabs, threeparttable}

\title{Neural Demand Estimation with Habit Formation and Rationality Constraints}
\author{Marta Grzeskiewicz\footnote{mg2192@cam.ac.uk. I am grateful to Rhys Williams for  valuable comments and suggestions.} \\ University of Cambridge}
\date{}

\begin{document}

\maketitle

\begin{abstract}
We develop a flexible neural demand system for continuous budget allocation that estimates budget shares on the simplex by minimizing KL divergence. Shares are produced via a softmax of a state-dependent preference scorer and disciplined with regularity penalties (monotonicity, Slutsky symmetry) to support coherent comparative statics and welfare without imposing a parametric utility form. State dependence enters through a habit stock defined as an exponentially weighted moving average of past consumption. Simulations recover elasticities and welfare accurately and show sizable gains when habit formation is present. In our empirical application using Dominick’s analgesics data, adding habit reduces out-of-sample error by ~33\%, reshapes substitution patterns, and increases CV losses from a 10\% ibuprofen price rise by about 15–16\% relative to a static model. The code is available at \url{https://github.com/martagrz/neural_demand_habit}.
\end{abstract}


\section{Introduction}

Understanding how households allocate expenditure across goods is central to evaluating inflation, taxation, and welfare policy. Demand systems deliver the workhorse mapping from prices and expenditure to budget shares and quantities, and—when disciplined by integrability—provide the basis for money-metric welfare analysis. The canonical continuous-allocation benchmark is the Almost Ideal Demand System (AIDS), which combines tractability with theory-based restrictions on substitution patterns \citep{deaton1980economics}. Yet modern applications increasingly confront three forces that strain classical specifications: (i) \emph{high-dimensional} price variation across many goods and attributes; (ii) \emph{state dependence} in consumption arising from habits, stockpiling, and reference dependence; and (iii) the need to compute welfare objects from models that must be both flexible and economically coherent.

This paper develops a neural econometric approach to demand estimation that treats consumer theory as a source of constrained semi-parametric discipline rather than as a fully parametric utility specification. The object we estimate is the conditional distribution of budget shares given the economic state. We model observed shares as compositional data on the probability simplex and estimate a \emph{neural demand system} by minimizing the Kullback--Leibler divergence between observed and predicted allocations. Predicted budget shares are generated by a softmax mapping of a flexible “preference scorer” that takes prices and expenditure as inputs and outputs per-good logits. This parameterization enforces adding-up by construction and directly targets the empirical object of interest—continuous budget allocation—without requiring a nested solution of a forward utility-maximization problem. In this sense, the estimator is a direct analogue of behavioural cloning in imitation learning: it learns the policy mapping from state to action under a likelihood-based loss, while remaining interpretable through the economic objects implied by the fitted demand system.

We study continuous within-category budget allocation across a small set of inside goods and focus on how state dependence changes elasticities and welfare derived from this mapping. Product-space discrete-choice models (e.g., BLP) are a natural complement when the object is substitution across differentiated UPCs; our goal is different: to provide a scalable, theory-disciplined share system for continuous allocation and welfare, and to show that even at this level, ignoring state can conflate segmentation with substitution.

Flexibility alone is not sufficient for welfare analysis. Welfare identification from demand relies on integrability conditions, most prominently Slutsky symmetry and curvature, that connect observed comparative statics to a well-behaved utility representation. We therefore introduce regularity penalties into the training objective that shrink the estimated system toward economically coherent comparative statics. In particular, we penalize violations of Slutsky symmetry and monotonicity, and we report diagnostic measures of near-integrability. The goal is not to claim that arbitrary neural predictors are rationalizable, but to provide a scalable estimator whose implied substitution patterns can be disciplined and transparently audited in applications where welfare is the target object.

A second motivation is that many consumption decisions are inherently dynamic. Static demand systems treat period-by-period choices as conditionally independent given contemporaneous prices and expenditure, abstracting from habit formation and other forms of state dependence. To capture path dependence without committing to a parametric habit-utility form, we augment the consumer state with a \emph{habit stock} constructed as an exponentially weighted moving average of past consumption. The preference scorer conditions on this habit stock nonparametrically, allowing the data to determine how state dependence interacts with prices and expenditure.

Introducing a habit stock raises an additional econometric issue: the \emph{habit decay parameter} governing the moving average controls the effective memory length of the state. With a flexible scorer, different decay values can generate habit inputs that are observationally similar over the support of the data, making the decay parameter weakly identified from within-sample fit. We therefore treat the decay rate as a structural hyperparameter and select it using a profile criterion; to reflect weak identification, we report an identified set of near-minimizers and summarize welfare conclusions over this set.

The paper proceeds in two parts. First, we validate the estimator in simulation across standard data-generating processes, where the ground truth for elasticities and compensating variation is known. This permits evaluation not only of predictive accuracy but also of structural recovery of own- and cross-price effects and welfare under exogenous price shocks. Second, we apply the framework to Dominick's Finest Foods scanner data \citep{kilts_dominicks_2018}, using the application primarily to demonstrate external validity of the econometric approach and to illustrate how state dependence and economic regularization can materially affect welfare conclusions relative to static benchmarks.

Our contributions are as follows. (1) We propose a neural demand estimator for continuous budget allocation on the simplex, estimated by KL minimization and designed to scale with dimensionality while preserving basic accounting structure by construction. (2) We incorporate consumer-theory discipline through regularity penalties and near-integrability diagnostics, enabling welfare analysis from a flexible demand model without imposing a parametric utility specification. (3) We extend the framework to state-dependent demand via a habit-augmented state and provide a practical profiling approach for the habit decay parameter, emphasizing robustness of welfare conclusions over its identified set and we show empirically how conditioning on the state separates structural substitution from persistent segmentation/state selection in scanner data. Together, these elements position the approach as a bridge between neural econometrics and constrained semi-parametric demand analysis, with welfare as the organizing application domain. 

\subsection{Literature}

Our positioning is closest to work that uses flexible (potentially high-dimensional) demand models while disciplining them with implications of consumer theory, and to recent econometric practice that treats theory as a form of regularization rather than an exact parametric specification. We view this as complementary to product-space discrete-choice demand (e.g., BLP) used for UPC-level differentiation: our object is continuous within-category \emph{budget allocation} on the simplex and the associated welfare functionals.

\subsubsection{Constrained Nonparametric Demand and Welfare}
A large literature studies demand systems that are flexible enough to fit rich substitution patterns while remaining consistent with (or close to) the integrability conditions that underlie welfare measurement. Classic parametric share systems such as AIDS \citep{deaton1980economics} and QUAIDS \citep{banks1997quadratic} deliver tractable elasticities and welfare but rely on restrictive functional forms that can be fragile when price responses are complex or high-dimensional \citep{blundell2004endogeneity}. Nonparametric and semiparametric approaches relax functional form assumptions while imposing shape restrictions derived from consumer theory. Early contributions include nonparametric tests of Slutsky symmetry \citep{lewbel1995consistent} and methods to test and impose symmetry in nonparametric share systems \citep{haag2009testing}. More recent work develops nonparametric estimators of demand under Slutsky-type inequality restrictions in settings with nonseparable heterogeneity \citep{blundell2013nonparametric}, and provides econometric treatments of inference under nonlinear shape restrictions \citep{HOROWITZ2017108}. This line of work is complemented by surveys and book-length treatments of shape constraints and rationality restrictions in economics \citep{johnson2018shape, chambers2016revealed}.

Our approach contributes to this agenda by combining (i) a geometry-respecting parameterization of budget shares on the simplex with (ii) regularity penalties that shrink estimates toward economically coherent comparative statics (e.g., Slutsky symmetry), and (iii) explicit diagnostic reporting of near-integrability to support welfare calculations. In spirit, this is related to ``shrinking toward theory'' estimators that trade off statistical fit and approximate satisfaction of theoretical restrictions \citep{fessler2019use}. Relative to classical nonparametric implementations of Slutsky restrictions, our framework scales naturally to many goods and permits state dependence through a low-dimensional habit state.

A complementary strand addresses the ``many prices'' problem using modern regularization while targeting welfare objects. \citet{chernozhukov2018double} develop estimators for average demand and associated bounds on exact consumer surplus when the price vector is high-dimensional, using debiased machine learning tools. While that framework focuses on average demand and surplus bounds under high-dimensional controls, we focus on individual-level budget shares (continuous allocation on the simplex), discipline the estimated system with consumer-theory regularity, and compute welfare objects from the estimated demand while assessing near-integrability via diagnostics.

Finally, recent work revisits what aspects of rationality are testable in fully nonparametric settings. In particular, \citet{gunsilius2025nonparametric} develop a fully nonparametric approach to testing Slutsky symmetry, highlighting both the conceptual and statistical challenges created by multidimensionality. Our paper is not about formal hypothesis testing of rationality restrictions; instead, we use these restrictions as economically motivated regularizers and report their empirical magnitude, aligning the welfare interpretation with the measured degree of near-integrability.

\subsubsection{State Dependence and Habit Formation}
Habit formation has a long theoretical tradition in demand analysis. Early foundations include
\citet{pollak1970habit} and \citet{spinnewyn1981rational}, and subsequent empirical work estimates
dynamic demand systems with state dependence under parametric habit or inertia specifications (e.g.,
\cite{browning2003shocks}; see also \cite{dube2010state, bronnenberg2012evolution} for evidence on
consumer inertia and persistent brand preferences in scanner data). Related demand-system evidence on
habits and heterogeneity in budget allocations is provided by \citet{browning2007habits}. Our contribution relative to this literature is to accommodate habit formation nonparametrically: the neural preference scorer conditions on a habit stock without restricting its functional role. This preserves the flexibility of the demand system while allowing the data to determine whether habit affects own- and cross-price responses and associated welfare measures.

\subsubsection{Machine Learning in Structural Econometrics}
A growing econometrics literature uses machine learning to flexibly model nuisance functions or
high-dimensional objects while retaining interpretable structural outputs; see, among many others,
\citet{bajari2015machine} and \citet{chernozhukov2018double}, and the survey in \citet{iskhakov2020machine}.
Related work also explores neural-network approaches to structural estimation in simulated or
algorithmic settings (e.g., \cite{wei2025estimating}). Our framework follows this spirit in a demand setting: the neural component provides flexible approximation, while consumer-theory implications enter as regularization and diagnostics for the derived elasticities and welfare objects.

Our treatment of the habit decay parameter $\delta$ connects to the partial identification literature \citep{tamer2010partial}. In flexible state-dependent demand models, different decay parameters can generate habit stocks that are highly collinear or observationally similar over the support of the data, making $\delta$ weakly identified from fit. We therefore use a profile criterion and report an identified set for $\delta$ in applications, emphasizing robustness of substantive conclusions over that set.

A closely related emerging strand uses neural networks to recover utility functions under revealed-preference discipline.
For example, \citet{grzeskiewicz2025pearl} propose a framework that combines revealed preference with inverse reinforcement learning to recover a concave neural utility representation (via an input-concave neural network) that rationalizes observed choices under a budget constraint. Our approach differs in both object and implementation: we estimate the conditional mean share system directly on the simplex via KL minimization (without a nested ``solve-the-agent'' loop), discipline the resulting comparative statics through regularization and diagnostics, and focus on state dependence and welfare robustness when the persistence parameter is weakly identified.

\subsubsection{Links to Imitation Learning}
Methodologically, our estimator can be viewed as direct estimation of the conditional distribution of budget shares given the state vector via KL minimization. This mirrors the logic of behavioural cloning in imitation learning \citep{pomerleau1991efficient}, which learns a policy by supervised learning from observed state--action pairs, rather than solving a nested forward optimization problem. In contrast, inverse reinforcement learning \citep{ng2000algorithms, ziebart2008maximum} targets the underlying objective (reward) and typically requires repeatedly solving a forward control problem during estimation (see also \cite{ho2016generative} for a modern adversarial formulation that connects IL and IRL). In our demand setting, KL minimization on the simplex provides a computationally direct route to predicting shares while still allowing recovery of economically relevant structural quantities (elasticities and welfare objects) from the fitted system, subject to the strength of the regularity diagnostics.

\section{A Neural Demand System with Habit Formation}
\label{sec:methodology}

\subsection{The Consumer's Decision Problem}
\label{sec:consumer}

At each period $t$, a household allocates total expenditure $y_t$ across $G$ goods at prices $p_t \in \mathbb{R}^G_{++}$, choosing budget shares $w_t \in \Delta^{G-1}$, where $w_{it} = p_{it} x_{it} / y_t$ and $\sum_i w_{it} = 1$. In the static formulation the household's decision depends only on contemporaneous prices and income. In the habit-augmented formulation, past consumption also influences current choices via a habit stock
\begin{equation}
  \bar{x}_t(\delta) = \delta\, \bar{x}_{t-1}(\delta) + (1-\delta)\, x_{t-1},
  \qquad \delta \in (0,1),
  \label{eq:habit_stock}
\end{equation}
an exponentially weighted moving average of past consumption with decay parameter $\delta$. The habit stock is a parsimonious state variable summarizing purchase history; in scanner data it can capture loyalty/habit persistence as well as inventory/stockpiling and other slow-moving demand shifters correlated with past purchases.\footnote{In the Dominick’s application, aggregate store–week purchases cannot distinguish preference-based habit from inventory dynamics (e.g., stockpiling during promotions). We therefore interpret $\bar x$ as a reduced-form state capturing persistence in allocations; the placebo (shuffled histories) and fixed-effect variants assess whether the predictive gains reflect genuine temporal structure rather than spurious additional covariates. Separating habit from inventory would require household panels or inventory proxies.} A household with a high habit stock for good $i$ demands more of good $i$ at any given $(p_t, y_t)$ than an otherwise identical household with a low habit stock.

The decay parameter $\delta$ controls the effective memory of the habit stock. A key observation is that $\delta$ is not identified from the within-sample estimation objective. For any two values $\delta$ and $\delta'$ producing habit stocks $\bar{x}(\delta)$ and $\bar{x}(\delta')$ that differ only by a smooth transformation, the neural preference scorer can reparameterise its hidden layers to produce identical predicted shares, leaving the KL objective unchanged. We therefore treat $\delta$ as a structural hyperparameter and select it via the profile criterion of Section~\ref{sec:delta_profile}.

The household's information set at period $t$ is
\begin{equation}
  s_t = \bigl(p_t,\; y_t,\; \bar{x}_t(\delta)\bigr).
  \label{eq:state}
\end{equation}
In the static formulation $\bar{x}_t$ is omitted and $s_t = (p_t, y_t)$.

Prices in the simulation follow $p_t = Z_t + \nu_t$, where $Z_t \in \mathbb{R}^G$ is a vector of cost shifters and $\nu_t \sim \mathcal{N}(0, 0.1 I)$. Income is drawn i.i.d.\ from $y_t \sim \mathcal{U}[1200, 2000]$.

\subsection{Linear Demand System}
\label{sec:linear_nds}

As a computationally tractable analogue to the full neural model, we estimate a linear demand system by KL-minimisation. The per-good preference score is linear in features:
\begin{equation}
  V_i(s; \theta) = \theta_i^\top \phi(s),
  \label{eq:linear_scorer}
\end{equation}
where $\phi : \mathcal{S} \to \mathbb{R}^K$ is a feature map and $\theta_i \in \mathbb{R}^K$ are good-specific parameters. Budget shares are recovered via softmax, and the model is estimated by minimising the batch-mean KL divergence with $\ell_2$ regularisation, updating $\theta$ by gradient ascent with decaying learning rate $\eta_t = \eta_0 / (1 + t/T_0)$, $\eta_0 = 0.05$, $T_0 = 1000$.

We consider three feature specifications, all using $\log p_i$, $\log^2 p_i$, and $\log y$ as base regressors.

\begin{enumerate}
  \item[(i)] \textit{Shared features (baseline).}
    Each good $g$ receives $[\log p_g,\; \log^2 p_g,\; \log y]$, forcing identical sensitivity profiles across goods.

  \item[(ii)] \textit{Good-specific features.}
    Each good $g$ receives all $G$ log-prices plus $\log^2 p_g$ and $\log y$, allowing heterogeneous cross-price sensitivities but introducing collinearity.

  \item[(iii)] \textit{Orthogonalised features with per-good intercepts.}
    The demeaned log-price matrix is QR-factored; each good receives a good-specific indicator, the $G$ orthogonal price factors, $\log^2 p_g$, and $\log y$:
    \[
      \phi_i(p, y) = \bigl[e_i,\; Q_{:,1},\; Q_{:,2},\; Q_{:,3},\;
      (\ln p_i)^2,\; \ln y\bigr],
    \]
    where $Q$ is the orthonormal factor of $\ln p - \overline{\ln p}$. 
\end{enumerate}

All three are estimated by projected gradient ascent on the KL objective with $L_2$ regularisation ($\lambda = 10^{-4}$).

\subsection{Neural Demand System}
\label{sec:nds}

\subsubsection{Network Architecture.}

The preference scorer $V_\psi(s)$ is a four-layer MLP with SiLU activations \citep{elfwing2018sigmoid}:
\[
\text{input} \;\xrightarrow{[d_s \to H]}\; \text{SiLU} \;\xrightarrow{[H \to H]}\; \text{SiLU} \;\xrightarrow{[H \to H/2]}\; \text{SiLU} \;\xrightarrow{[H/2 \to G]}\; \text{logits}
\]
where $d_s$ is the input dimension ($G+1$ static, $2G+1$ habit-augmented), $H=256$, and $G=3$. SiLU is preferred over ReLU-family activations for its smooth gradient, which provides implicit regularisation suited to recovering concave preference surfaces.

Budget shares are recovered via softmax:
\begin{equation}
  \hat{w}_i(s;\, \psi)
  = \frac{\exp\bigl([V_\psi(s)]_i\bigr)}
         {\sum_{j=1}^G \exp\bigl([V_\psi(s)]_j\bigr)},
  \label{eq:softmax_shares}
\end{equation}
satisfying adding-up by construction. The scale of $V_\psi$ is not separately identified from the output layer weights, so we fix the softmax temperature at unity without loss of generality, analogously to the standard logit scale normalisation. Weights are initialised with Kaiming normal; the final layer uses Xavier uniform with gain 0.1 to prevent large-logit saturation.

The \emph{static} model takes $s = [\ln p_1, \ldots, \ln p_G, \ln y] \in \mathbb{R}^{G+1}$. The \emph{habit-augmented} model takes
\begin{equation}
  s^{\text{hab}} = [\ln p_1,\, \ldots,\, \ln p_G,\, \ln y,\,
  \bar{x}_1(\hat\delta),\, \ldots,\, \bar{x}_G(\hat\delta)] \in \mathbb{R}^{2G+1},
  \label{eq:hab_input}
\end{equation}
where $\bar{x}(\hat\delta)$ is pre-computed once at the profile-selected $\hat\delta$ before training begins.

\subsubsection{Training Objective.}

The model is estimated by minimising:
\begin{equation}
  \mathcal{L}(\psi)
  = \underbrace{D_{\mathrm{KL}}\!\left(w^{\mathrm{obs}}
      \,\|\, \hat{w}\right)}_{\text{fit}}
  + \lambda_{\mathrm{mono}} \underbrace{\frac{1}{G}\sum_i
      \mathbb{E}\!\left[\max\!\left(0,\;
      \tfrac{\partial \hat{w}_i}{\partial \ln p_i}\right)\right]
    }_{\text{monotonicity}}
  + \lambda_{\mathrm{slut}} \underbrace{\mathbb{E}\!\left[
      \|S - S^{\top}\|_F^2\right]}_{\text{Slutsky symmetry}},
  \label{eq:loss}
\end{equation}
where $D_{\mathrm{KL}}$ is the batch-mean KL divergence between observed and predicted budget shares. KL divergence is the natural loss for compositional data on the probability simplex and corresponds to maximum likelihood estimation of a softmax model.

The monotonicity penalty $\mathcal{L}_{\mathrm{mono}} = \mathbb{E}[\max(0, \partial \hat{w}_i / \partial \ln p_i)]$ penalises upward-sloping Marshallian demands, ruling out Giffen behaviour. The Slutsky symmetry penalty penalises asymmetry in the Slutsky substitution matrix $S$, whose elements are computed from predicted shares via the Slutsky equation:
\begin{equation}
  S_{ij} = \frac{\partial \hat{w}_i}{\partial \ln p_j} + \hat{w}_j \frac{\partial \hat{w}_i}{\partial \ln y},
  \label{eq:slutsky}
\end{equation}
with all derivatives computed by automatic differentiation. Symmetry of $S$ --- $S_{ij} = S_{ji}$ --- is a necessary condition of rational utility maximisation. This is the correct Slutsky symmetry condition on compensated cross-price effects; the Marshallian Jacobian $J_{ij} = \partial \hat{w}_i / \partial \ln p_j$ does not satisfy symmetry in general and is not penalised. The Slutsky penalty is activated after epoch 1,000 to allow the KL loss to converge first. Hyperparameters: $\lambda_{\mathrm{mono}} = 0.20$, $\lambda_{\mathrm{slut}} = 0.10$.

\subsubsection{Optimisation.}

All models are optimised with Adam \citep{kingma2014adam} at learning rate $5 \times 10^{-4}$, $L_2$ weight decay $10^{-5}$, and gradient-norm clipping at 1.0. The simulation pipeline trains for 10{,}000 epochs with mini-batches of 256; the Dominick's pipeline uses 3{,}000 epochs (static) and 4{,}000 epochs (habit-augmented) with mini-batches of 512. The best checkpoint by full-dataset KL, evaluated every 50 epochs, is retained.

\subsection{Neural Demand System with Control Function Instruments}
\label{sec:neural_iv}

Observed store-week prices may be correlated with unobserved demand shocks (e.g., promotions), so a
demand model that conditions directly on prices can be endogeneity biased. We address this using a
control function \citep{rivers1988limited, blundell1993we}. First-stage diagnostics are reported in
Appendix~\ref{sec:first_stage}.

For each good $g$, we obtain a residual from a first-stage projection of log price on instruments (on
the same scale as the second stage):
\begin{equation}
  \ln p_{git} = \tilde Z_{it}^\top \pi_g + v_{git}, \qquad
  \hat v_{git} \equiv \ln p_{git}-\tilde Z_{it}^\top \hat\pi_g,
  \label{eq:first_stage_short}
\end{equation}
where $\tilde Z_{it}$ stacks $\ln Z_{it}$ (and any included first-stage controls).\footnote{This scale
alignment is immaterial when instruments are already in log form (as in the Dominick's Hausman
instruments) but matters in simulation designs where prices are generated additively in levels,
$p=Z+\nu+\xi$, so that $\ln p$ is approximately affine in $\ln Z$ on the observed support.}

In the second stage we augment the neural demand input with the residual vector
$\hat v_{it}=(\hat v_{1it},\ldots,\hat v_{Git})^\top$:
\begin{equation}
  s^{\mathrm{CF}}_{it} =
  \bigl[\ln p_{it},\; \ln y_{it},\; \bar x_{it}(\hat\delta),\; \hat v_{it}\bigr]\in\mathbb R^{3G+1},
  \label{eq:cf_input_short}
\end{equation}
and estimate the network by minimizing \eqref{eq:loss}. In the Dominick's application, we use
Hausman-style ``other-store same-week'' prices as instruments because they are available at the same
level of aggregation as the share system and naturally capture common cost variation across the
retail network. Product-characteristics (``differentiation'') instruments are standard in UPC-level BLP
applications but are not directly available in our three-good aggregation, which abstracts from
within-ingredient characteristics and manufacturer ownership.

For counterfactual prediction we evaluate the fitted network at $\hat v_{it}=\mathbf 0$,
\[
\hat w^{\mathrm{CF}}(p,y,\bar x)\equiv \hat w(\ln p,\ln y,\bar x,\mathbf 0),
\]
which recovers the demand mapping associated with exogenous price variation under standard
control-function conditions. Regularity penalties in \eqref{eq:loss} are computed at $\hat v=\mathbf 0$,
so symmetry and monotonicity diagnostics pertain to the purged (causal) mapping rather than to an
endogeneity-contaminated fit.

\subsection{Profiled Estimation of the Habit-Decay Parameter}
\label{sec:delta_profile}

The habit-decay parameter $\delta$ has a clear structural interpretation. It determines how rapidly past consumption is discounted, but is not identified from the within-sample KL objective. For any $\delta$, the neural preference scorer can remap $\bar{x}(\delta)$ through its hidden layers; two values of $\delta$ producing habit stocks that differ by a smooth monotone transformation are observationally equivalent from the network's perspective, so the KL loss is flat across such pairs. Gradient descent through the network therefore converges to an attractor determined by initialisation rather than the data-generating process.

We recover $\delta$ by a profile criterion external to the network's training. Let $\mathcal{D}=\{\delta_1,\ldots,\delta_K\}\subset(0,1)$ be a fine grid. For each $\delta_k$, pre-compute $\bar{x}(\delta_k)$, train the network to convergence, and evaluate out-of-sample KL:
\begin{equation}
  \hat{\psi}(\delta_k)
  \in \arg\min_{\psi}\; \mathcal{L}\!\left(\psi;\delta_k\right),
  \label{eq:inner_opt}
\end{equation}
\begin{equation}
  \widehat{L}_{\text{test}}(\delta_k)
  = D_{\mathrm{KL}}\!\left(w^{\mathrm{obs}}_{\text{test}} \,\|\, \hat{w}_{\hat{\psi}(\delta_k)}(\cdot;\delta_k)\right).
  \label{eq:profile_loss}
\end{equation}
The profile-selected value is $\hat{\delta} \in \arg\min_{\delta_k} \widehat{L}_{\text{test}}(\delta_k)$. Because the profile loss is typically flat over a wide range, we report an identified set:
\begin{equation}
  \widehat{\mathcal{D}}
  = \left\{\delta_k:\;
  \widehat{L}_{\text{test}}(\delta_k)\le
  \min_{\delta\in\mathcal{D}}\widehat{L}_{\text{test}}(\delta) + c\cdot \widehat{\mathrm{SE}}\right\},
  \label{eq:identified_set}
\end{equation}
where $\widehat{\mathrm{SE}}$ is computed by block bootstrap over stores and $c\in\{1,2\}$ is a one- or two-standard-error rule. Welfare objects are reported as bounds over $\widehat{\mathcal{D}}$, ensuring policy conclusions do not depend on a point estimate of a partially identified parameter.

\subsection{Window Demand System}
\label{sec:window}

As a non-parametric alternative to the EWMA habit specification, we estimate a Window model conditioning on the $L$ most recent periods of (log price, log quantity) pairs:
\begin{equation}
  s_t = [\log p_t,\; \log y_t,\; \log p_{t-1},\; \log q_{t-1},\; \ldots,\; \log p_{t-L},\; \log q_{t-L}] \in \mathbb{R}^{G + 1 + 2GL}.
\end{equation}
Missing history at store boundaries is filled with global column-wise means. We use $L = 4$ lags. The Window model uses the same MLP and training procedure as the neural demand system; there is no $\delta$ to select.

\subsection{Identification}
\label{sec:identification}

The primitive observables are $(w_t,p_t,y_t)$ and a habit stock $\bar x_t(\delta)$ constructed from
lagged consumption. Fixing $\delta$, define the state
$s_t(\delta)\equiv(p_t,y_t,\bar x_t(\delta))$ (or $(p_t,y_t)$ in the static model). Our target object is
the conditional mean share system
\begin{equation}
m_\delta(s)\equiv \mathbb E[w_t\mid s_t(\delta)=s], \qquad w_t\in\Delta^{G-1}.
\label{eq:m_def}
\end{equation}
The neural demand system is a flexible approximation to $m_\delta(\cdot)$ with adding-up by
construction and economic regularization.

\subsubsection{Demand for fixed $\delta$.}
Under standard support conditions, $m_\delta(\cdot)$ is identified in the usual nonparametric sense
\citep{matzkin2003nonparametric}. The latent scorer $V_\psi$ is not separately identified (logit-scale
normalization); economically meaningful objects are the implied shares and their derivatives.

\subsubsection{Welfare functionals.}
Given an estimated share system $\hat w$, compensating variation is computed as the line integral
\begin{equation}
\widehat{\mathrm{CV}}
\approx
-\int_{p_0}^{p_1}\hat q(p)\,dp,
\qquad
\hat q_i(p)=\frac{\hat w_i(p,y,\bar x)\,y}{p_i}.
\label{eq:cv_line_integral}
\end{equation}
When demand is integrable, CV is a functional of demand and does not require cardinal utility
\citep{hausman1981exact,vartia1983efficient}. In our framework, adding-up holds exactly and the
remaining integrability conditions (homogeneity, symmetry, negativity) are treated as regularization:
we penalize symmetry/monotonicity violations and report near-integrability diagnostics. In the
application, $y_{it}$ is within-category expenditure, so $\widehat{\mathrm{CV}}$ is measured in
category-expenditure units.

\subsubsection{Habit effects and $\delta$.}
For fixed $\delta$, habit effects are derivatives with respect to the state,
\begin{equation}
H_{ij}(s;\delta)\equiv \frac{\partial m_{\delta,i}(s)}{\partial \bar x_j},
\label{eq:habit_effect_def}
\end{equation}
identified wherever $\bar x(\delta)$ varies locally conditional on $(p,y)$. In the panel application we
use fixed-effect variants as a robustness check against persistent store heterogeneity.

Finally, the decay parameter $\delta$ is generally weakly (often set) identified from fit in flexible
state-dependent models because $\bar x(\delta)$ processes are highly collinear over observed support.
We therefore profile over $\delta$ using out-of-sample KL loss and report an identified set
$\widehat{\mathcal D}$ of near-minimizers; substantive results are reported at $\hat\delta$ and/or over
$\widehat{\mathcal D}$ \citep{tamer2010partial}.

\subsection{Traditional Demand System Benchmarks}
\label{sec:benchmarks}

\subsubsection{Linear Approximate AIDS.}
The AIDS of \citet{deaton1980economics} specifies
\begin{equation}
  w_i = \alpha_i + \sum_{j} \gamma_{ij} \ln p_j
        + \beta_i \ln\!\left(\frac{y}{P}\right),
  \label{eq:aids}
\end{equation}
estimated via OLS. AIDS has three limitations: (i) income enters every share equation by construction, preventing zero income effects even when the true DGP is quasilinear; (ii) the log-linear functional form cannot represent subsistence constraints or kinked demand; and (iii) it is inherently static.

\subsubsection{Quadratic AIDS (QUAIDS).}
\citet{banks1997quadratic} extend AIDS by adding a quadratic log-income term to each share equation:
\begin{equation}
  w_i = \alpha_i + \sum_{j} \gamma_{ij} \ln p_j
        + \beta_i \ln\!\left(\frac{y}{a(p)}\right)
        + \frac{\lambda_i}{b(p)} \left[\ln\!\left(\frac{y}{a(p)}\right)\right]^2,
  \label{eq:quaids}
\end{equation}
where $a(p) = \alpha_0 + \sum_i \alpha_i \ln p_i + \frac{1}{2}\sum_i\sum_j \gamma_{ij} \ln p_i \ln p_j$ is the translog price index and $b(p) = \prod_i p_i^{\beta_i}$ is a Cobb--Douglas price aggregator. The adding-up restrictions require $\sum_i \alpha_i = 1$, $\sum_i \gamma_{ij} = \sum_j \gamma_{ij} = 0$, $\sum_i \beta_i = 0$, and $\sum_i \lambda_i = 0$; Slutsky symmetry requires $\gamma_{ij} = \gamma_{ji}$. QUAIDS is estimated by iterated linear least squares \citep{blundell1993we}, imposing adding-up and symmetry restrictions. It accommodates non-linear Engel curves and is the natural parametric benchmark for testing whether the neural preference scorer's flexibility over income effects is empirically important. QUAIDS shares AIDS's static limitation and its inability to represent subsistence minima or kinked demand.

\subsubsection{Nonparametric Demand.}
\citet{Blundelletal2012} propose a nonparametric estimator for Engel curves and demand functions that imposes the Slutsky negativity and symmetry restrictions as shape constraints rather than parametric restrictions. We implement the shape-constrained series estimator of \citet{Blundelletal2012} as follows. For each good $i$, we estimate
\begin{equation}
  w_i = m_i(p, y) + \varepsilon_i,
  \label{eq:np_demand}
\end{equation}
where $m_i : \mathbb{R}^G_{++} \times \mathbb{R}_{++} \to [0,1]$ is approximated by a tensor-product B-spline basis with knots chosen by cross-validation. Adding-up is imposed by estimating only $G-1$ equations and recovering the $G$th by $w_G = 1 - \sum_{i<G} \hat{w}_i$. Slutsky symmetry is imposed as a linear constraint on the B-spline coefficients following the approach of \citet{Blundelletal2012}: the matrix of cross-partial derivatives of the spline with respect to log-prices, evaluated at a grid of representative $(p,y)$ points, is constrained to be symmetric. This constitutes the most flexible benchmark in our comparison that is not neural: it imposes no functional-form restrictions on income effects or substitution patterns, while maintaining the same integrability constraints as the neural demand system.

The nonparametric estimator is static by construction and does not accommodate habit formation. In the habit DGP and the Dominick's application it therefore serves as a ceiling on what a static, nonparametric, correctly-constrained estimator can achieve --- any gains of the neural demand system over it on static DGPs reflect approximation advantages of the neural architecture, while gains on the habit DGP reflect the structural value of habit-state augmentation.

\subsubsection{BLP Logit-IV.}
\label{sec:blp}
Following \citet{berry1995automobile}, we implement a full-price 2SLS BLP specification:
\begin{equation}
  \delta_g = \alpha_{g0} + \sum_j \alpha_{gj} p_j,
  \label{eq:blp}
\end{equation}
instrumented with $G$ Hausman instruments (mean price across other stores in the same week). BLP is generally not an appropriate structural comparator for the simulation DGPs, which generate continuous budget allocations without an outside option. BLP is therefore included in all predictive accuracy and welfare comparisons throughout the paper, but excluded from elasticity tables.\footnote{The exclusion reflects a structural distinction
rather than a performance judgement. Every other benchmark estimates continuous budget
allocation on the simplex; BLP estimates discrete choice probabilities with an outside
option. In the simulation, the DGP generates continuous allocations with no outside
option by construction, so the logit IIA assumption and outside-option normalisation
that underlie BLP's elasticity formula are violated by design and the implied
elasticities are structurally uninterpretable. In the Dominick's application the true
DGP is unknown, but the same concern applies: BLP's ibuprofen own-price elasticity
is an outlier driven by the logit IIA constraint binding in a market where
acetaminophen holds a 50\% expenditure share, not by a genuine difference in price
sensitivity. Predictive RMSE and compensating variation do not rely on the logit
functional form and remain informative comparators regardless of the DGP. }

\section{Simulation}
\label{sec:simulation}

We validate all models via a synthetic simulation with known ground truth before applying the framework to Dominick's data. Simulation permits exact evaluation of structural recovery --- elasticities and compensating variation --- quantities unobservable in real data.

Models evaluated: (1) LA-AIDS; (2) QUAIDS; (3) Nonparametric NDS (shape-constrained B-spline); (4) BLP (IV); (5) Linear NDS in three feature variants (Shared, Good-Specific, Orthogonalised); (6) Neural NDS (static, no IV); (7) Neural NDS-CF (static, control function IV); (8) Habit-Augmented NDS ($\delta$ selected by profile criterion, no IV); (9) Habit-Augmented NDS-CF ($\delta$ selected by profile criterion, control function IV). In the simulation, the true DGP satisfies the instrument exclusion restriction by construction (prices are generated as $p = Z + \nu$ with $Z$ excluded from the utility function), so the NDS-CF and NDS estimates should coincide in population; any divergence reflects finite-sample bias from the control function correction and serves as a specification check.

\subsection{Data Generating Processes}
\label{sec:dgps}

This subsection describes the simulation environments used to evaluate the estimator. The goal is
twofold. First, we study structural recovery in benchmark static models spanning smooth substitution
(CES), nonhomotheticity (Stone--Geary), and corner-like behavior (Leontief), as well as a quasilinear
environment with zero income effects on a subset of goods. Second, we study two settings that are
central to the empirical motivation: (i) price endogeneity and (ii) state dependence through habit
formation. In each design, we generate synthetic budget shares from a known utility model, train the
neural demand system exactly as in the empirical pipeline, and evaluate recovery of (a) budget shares,
(b) own- and cross-price elasticities, and (c) compensating variation under counterfactual price
changes. Unless otherwise noted, we set $G=3$ and normalize total expenditure to $y_t$.

\subsubsection{CES}

The CES environment provides a smooth, globally well-behaved demand system with closed-form
Marshallian shares. Preferences are
\begin{equation}
  U(x)=\Bigl(\sum_{i=1}^G \alpha_i x_i^{\rho}\Bigr)^{1/\rho},
  \label{eq:ces_utility}
\end{equation}
with $\alpha=(0.4,0.4,0.2)$ and $\rho=0.45$, implying elasticity of substitution
$\sigma = 1/(1-\rho)\approx 1.82$. For given prices $p$ and expenditure $y$, Marshallian budget shares
satisfy
\begin{equation}
  w_i^*(p)=
  \frac{\alpha_i^{\sigma}p_i^{1-\sigma}}{\sum_{j=1}^G \alpha_j^{\sigma}p_j^{1-\sigma}},
  \label{eq:ces_shares}
\end{equation}
and quantities are $x_i^*(p,y)=w_i^*(p)\,y/p_i$. This design isolates approximation and differentiation
accuracy in a setting where the demand system is smooth and fully integrable.

\subsubsection{Quasilinear}

To stress-test Engel curve flexibility, we consider a quasilinear utility with a numeraire good $x_0$:
\[
U(x)=x_0 + a_1\log(x_1+1)+a_2\log(x_2+1),
\]
with $(a_1,a_2)=(1.5,0.8)$ and $G=3$ (so goods 1--2 are non-numeraire and good 3 plays the role of
$x_0$ in the share system). Quasilinearity implies zero income effects on goods 1 and 2 over the
relevant region, a feature that standard share systems can struggle to approximate unless explicitly
parameterized to allow such restrictions. This benchmark is therefore informative about whether a
flexible share mapping can recover the correct income-response structure without parametric guidance.

\subsubsection{Leontief}

To generate sharp substitution patterns, we consider Leontief (perfect complements) preferences
\[
U(x)=\min\{x_0/a_0,\,x_1/a_1,\,x_2/a_2\},
\]
with $a=(1.0,0.8,1.5)$. Demands exhibit essentially zero substitution elasticity and can generate
kinks in the indirect utility, providing a stringent test for gradient-based recovery of elasticities and
for the stability of automatic differentiation in the neighborhood of non-smooth behavior. In practice,
we generate demands using the implied fixed-proportion consumption bundle and compute shares from
$p_ix_i/y$.

\subsubsection{Stone--Geary}

To mirror empirically relevant nonhomotheticities, we simulate from Stone--Geary preferences
\[
U(x)=\prod_{i=1}^G (x_i-\gamma_i)^{\alpha_i},
\]
with $\alpha=(0.5,0.3,0.2)$ and $\gamma=(50,30,20)$. This yields the linear expenditure system (LES)
with closed-form Marshallian demands:
\begin{equation}
  p_i x_i^* = p_i\gamma_i + \alpha_i\Bigl(y-\sum_{j=1}^G p_j\gamma_j\Bigr).
  \label{eq:les}
\end{equation}
Subsistence requirements induce nonlinear Engel curves and strong interactions between prices and
effective discretionary income, making this environment a natural bridge to the empirical scanner-data
application.

\subsubsection{CES with endogenous prices}

To evaluate the control-function correction in a setting with known ground truth, we simulate a CES
share system with price endogeneity generated by correlation between demand shocks and prices.
Shares follow the CES equation \eqref{eq:ces_shares} with time-varying taste shifters:
\begin{equation}
  w_{jt} =
  \frac{\alpha_{jt}^{\sigma} p_{jt}^{1-\sigma}}{\sum_{k=1}^G \alpha_{kt}^{\sigma} p_{kt}^{1-\sigma}},
  \qquad
  \alpha_{jt}=\bar\alpha_j\exp(\xi_{jt}),
  \label{eq:endog_ces_shares}
\end{equation}
where $\bar\alpha=(0.4,0.4,0.2)$ and $\xi_{jt}$ is a structural demand shock observed by the consumer
but not by the econometrician. Prices are generated from an instrument $Z_{jt}$, an idiosyncratic
supply shock $\nu_{jt}$, and the demand shock $\xi_{jt}$:
\begin{equation}
  p_{jt}=\max\{Z_{jt}+\nu_{jt}+\xi_{jt},\,0.1\}.
  \label{eq:endog_price}
\end{equation}
We draw $Z_{jt}\sim\mathcal U(1,5)$, $\nu_{jt}\sim\mathcal N(0,0.1^2)$, and
$\xi_{jt}\sim\mathcal N(0,0.5^2)$. Since $\xi_{jt}$ enters both \eqref{eq:endog_ces_shares} and
\eqref{eq:endog_price}, prices are positively correlated with unobserved demand determinants,
generating simultaneity bias. This design allows a direct comparison between (i) an uncorrected
neural demand system and (ii) the control-function variant, benchmarked against the known causal
demand mapping obtained by holding $\xi_{jt}$ fixed.

\subsubsection{Habit formation}

Finally, we simulate a habit-formation environment to evaluate whether augmenting the state with a
habit stock recovers the dynamic structure and improves welfare recovery under path dependence.
Preferences are CES-in-habits:
\begin{equation}
  U_t(x;\bar x_t)=
  \Bigl(\sum_{i=1}^G \alpha_i\bigl(x_i-\theta \bar x_{ti}\bigr)^{\rho}\Bigr)^{1/\rho},
  \label{eq:habit_utility}
\end{equation}
with $\alpha=(0.4,0.4,0.2)$, $\rho=0.45$, and habit intensity $\theta=0.3$. The habit stock evolves
as an exponentially weighted moving average,
\[
\bar x_{t}(\delta)=\delta \bar x_{t-1}(\delta)+(1-\delta)x_{t-1},
\]
with $\delta_{\mathrm{DGP}}=0.7$. Because $\bar x_t$ enters utility, identical contemporaneous
$(p_t,y_t)$ can generate different choices depending on consumption history, producing systematic
variation that static models must treat as noise. In this design we evaluate (i) the out-of-sample gains
from habit augmentation, (ii) recovery of the habit-induced comparative statics, and (iii) the extent to
which welfare conclusions are robust to weak identification of $\delta$ as characterized by the profile
criterion and the identified set $\widehat{\mathcal D}$.

\subsection{Experimental Design}
\label{sec:exp_design}

We generate $N = 800$ observations as follows. Instrument vectors are drawn 
$Z \sim \mathcal{U}[1,5]^G$ independently across observations, representing exogenous 
cost shifters that satisfy the exclusion restriction by construction. Prices are 
generated as $p = Z + \nu$, $\nu \sim \mathcal{N}(0, 0.1\,I)$, where $\nu$ is an 
idiosyncratic supply shock uncorrelated with the utility function; the small noise 
variance ($\sigma = 0.1$) ensures prices are close to their instrument values while 
preserving the price variation needed for elasticity identification. In the endogenous 
CES DGP (Section~\ref{sec:dgps}), prices are additionally shifted by the structural 
demand shock $\xi$, as described there. Income is drawn independently from 
$y \sim \mathcal{U}[1200, 2000]$, a range chosen to ensure that all goods have 
positive expenditure shares across the full price support under each DGP, and in 
particular, that the Stone--Geary subsistence requirements $\gamma = [50, 30, 20]$ are 
comfortably satisfied at all income realisations.

The sample size of $N = 800$ is deliberately modest. It is chosen to represent a 
realistic scanner-data cell size --- comparable to the number of store--week 
observations available for a single product category in a medium-sized retail panel 
--- rather than a large-$N$ asymptotic regime where all models would eventually 
converge. Under the CES DGP with $\alpha = [0.4, 0.4, 0.2]$ and $\rho = 0.45$, the 
resulting mean training budget shares are: Food 0.432, Fuel 0.438, Other 0.130. The 
asymmetry between Food/Fuel and Other reflects the lower preference weight 
$\alpha_2 = 0.2$ and is intentional: it creates a setting where models must accurately 
represent a minority good with a systematically lower share, which is a known 
difficulty for log-linear systems such as LA-AIDS that impose symmetric functional 
forms across goods.

Each DGP is evaluated across 5 independent seeds, with fresh draws of $(Z, \nu, y)$ 
at each seed; all models are re-estimated from scratch at each seed. Reported means 
and standard errors are computed across seeds. This design separates estimation 
variance (across seeds) from model bias (systematic departures from ground truth), 
allowing the tables to distinguish models that are consistently wrong from models 
that are noisy but unbiased.

\subsubsection{Price Shock and Evaluation Design.}
Models are trained on prices drawn from $p \sim Z + \nu$, $Z \sim \mathcal{U}[1,5]^G$, 
and evaluated on a counterfactual test set constructed by applying a 20\% shock to the 
Good-1 (Fuel) price: $p_1 \to 1.2\, p_1$, with all other prices and income held at 
their training-distribution values. Predictive accuracy is therefore measured at prices 
outside the training support rather than on a held-out draw from the same distribution. 
This design serves two purposes. First, it tests whether models have learned the 
structural shape of the demand system --- the slope and curvature of the demand curve 
--- well enough to extrapolate to an unseen price regime, rather than merely 
interpolating within the observed price range. A model that memorises the training 
distribution without recovering the underlying preference structure will fail here even 
if it fits the training data well. Second, since compensating variation for the same 
shock is the welfare object of interest, post-shock RMSE functions as a direct 
reliability check for the welfare results: a model that cannot accurately predict 
demand at $p_1^{\text{post}}$ will produce unreliable compensating variation for the 
transition from $p_1^{\text{pre}}$ to $p_1^{\text{post}}$.

\subsubsection{Evaluation Metrics.}

Three metrics are computed for each model and DGP, each targeting a distinct aspect of 
structural recovery.

\begin{enumerate}

  \item[(i)] \textit{Post-shock predictive RMSE and MAE.} Root mean squared error and 
  mean absolute error between predicted and observed budget shares on the post-shock 
  test set, averaged across goods and observations:
  \[
    \mathrm{RMSE} = \sqrt{\frac{1}{NG}\sum_{t=1}^{N}\sum_{i=1}^{G}
    \bigl(\hat{w}_{it} - w_{it}^{\mathrm{obs}}\bigr)^2}.
  \]
  As discussed above, these metrics are evaluated at post-shock prices rather than on a 
  held-out draw from the training distribution, so they measure counterfactual 
  extrapolation accuracy rather than in-distribution fit. A model with a correctly 
  recovered demand system will generalise to the new price regime; a model that has 
  overfit the training price cloud without learning the underlying preference structure 
  will not.

  \item[(ii)] \textit{Quantity elasticities.} The own-price quantity 
  elasticity for good $i$ is defined as
  \begin{equation}
    \hat{\varepsilon}_{ii} = \frac{\partial \hat{w}_i / \hat{w}_i}
    {\partial p_i / p_i} - 1,
    \label{eq:elasticity}
  \end{equation}
  and computed numerically via symmetric finite difference with step $h = 10^{-4}$,
  evaluated at mean post-shock prices and income. The $-1$ adjustment converts the 
  budget-share elasticity to a quantity elasticity under the assumption that total 
  expenditure $y$ is held fixed: since $w_i = p_i q_i / y$, a 1\% increase in $p_i$ 
  holding $y$ constant reduces $q_i$ by $1 + \hat\varepsilon_{ii}^w$ percent, where 
  $\hat\varepsilon_{ii}^w = \partial \ln \hat{w}_i / \partial \ln p_i$ is the 
  share elasticity. Cross-price elasticities are computed analogously and reported in 
  the heatmap figures. For the CES DGP, ground-truth elasticities are available in 
  closed form and serve as the structural recovery benchmark; for other DGPs, 
  elasticities are reported relative to the model consensus rather than a known truth.

  \item[(iii)] \textit{Compensating variation.} The compensating variation for a price 
  change from $p_0$ to $p_1$ is approximated via Riemann sum along a 100-step linear 
  price path:
  \begin{equation}
    \widehat{\mathrm{CV}} \approx -\int_{p_0}^{p_1} q(p)\, \mathrm{d}p,
    \quad q_i(p) = \frac{\hat{w}_i(p,y)\, y}{p_i},
    \label{eq:cv}
  \end{equation}
  where income $y$ is held fixed at its pre-shock mean throughout the integration. The 
  line integral is taken along the direct path $p(\tau) = p_0 + \tau(p_1 - p_0)$, 
  $\tau \in [0,1]$; path independence under Slutsky symmetry means the choice of path 
  does not affect the result for integrable demand systems, and provides a mild 
  specification check for models where symmetry is only approximately satisfied. For 
  the CES DGP, the true compensating variation is available in closed form via the 
  expenditure function, permitting exact evaluation of welfare recovery error. For 
  models that do not impose integrability, including the neural demand system, for 
  which Slutsky symmetry is penalised but not exactly enforced, the line integral 
  provides a consistent welfare approximation whose accuracy improves as near-integrability 
  diagnostics improve.

\end{enumerate}

\subsection{Simulation Results}
\label{sec:results}

\subsubsection{CES Ground Truth: Predictive Accuracy, Elasticities and Welfare}
\label{sec:sim_ces}

\paragraph{Predictive Accuracy.}

\begin{table}[htbp]
  \centering
  \caption{Post-Shock RMSE: CES DGP (mean $\pm$ SE, 5 runs)}
  \label{tab:sim_dgp_ces}
  \begin{threeparttable}
  \begin{tabular}{lc}
    \toprule
    \textbf{Model} & \textbf{CES} \\
    \midrule
    LA-AIDS & $0.00979 \pm 0.00010$ \\
    BLP (IV) & $0.02207 \pm 0.00025$ \\
    QUAIDS & $0.00978 \pm 0.00012$ \\
    Series Estm. & $0.00302 \pm 0.00006$ \\
    LD (Shared) & $0.15411 \pm 0.00030$ \\
    LD (GoodSpec) & $0.15532 \pm 0.00040$ \\
    LD (Orth) & $0.02398 \pm 0.00043$ \\
    ND (static) & $0.00063 \pm 0.00004$ \\
    ND (habit) & $0.00109 \pm 0.00010$ \\
    ND (CF) & $0.00075 \pm 0.00002$ \\
    ND (habit, CF) & $0.00080 \pm 0.00003$ \\
    \bottomrule
  \end{tabular}
  \begin{tablenotes}\small
    \item Post-shock: 20\% increase in price of good 1. LD: Linear Demand; ND: Neural Demand.
  \end{tablenotes}
  \end{threeparttable}
\end{table}

\begin{figure}[htbp]
  \centering
  \includegraphics[width=0.75\linewidth]{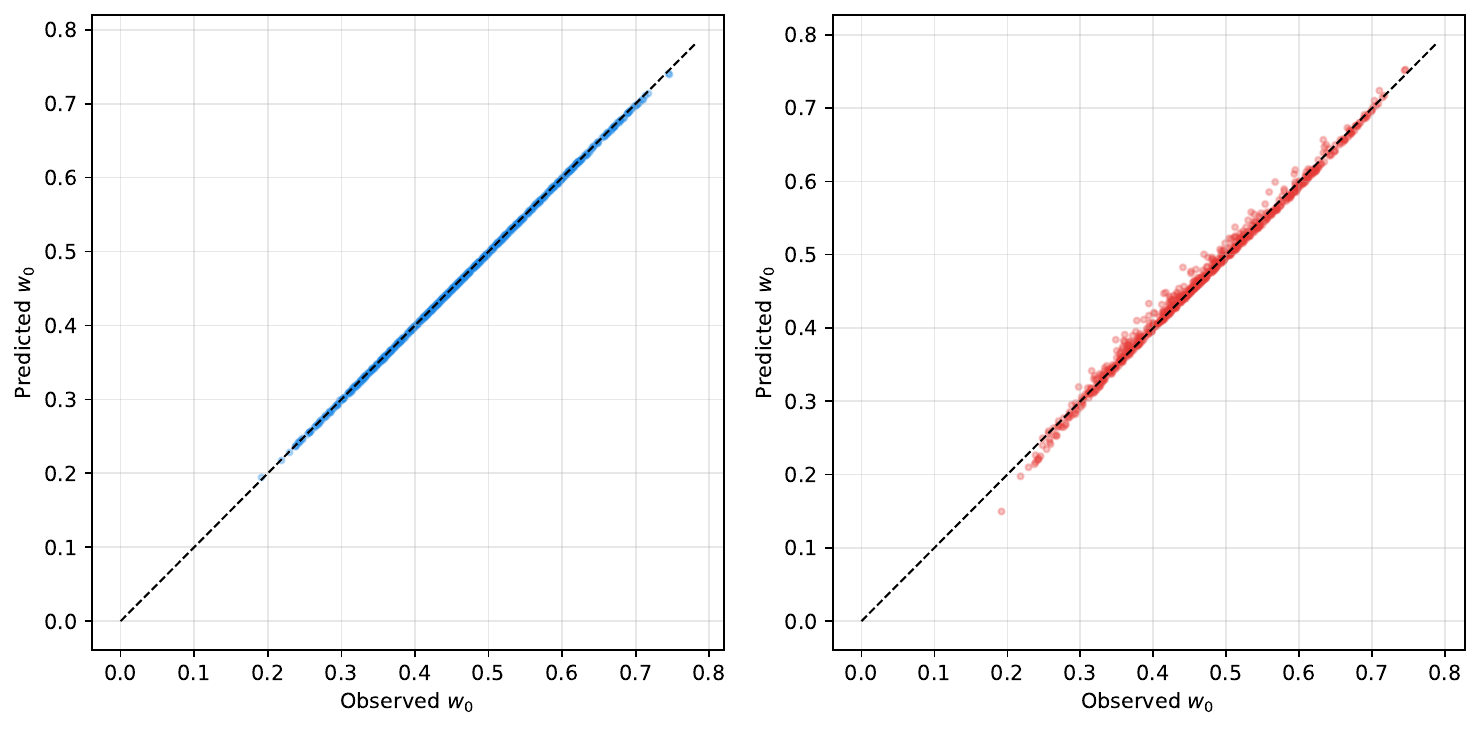}
  \caption{Observed vs.\ predicted budget share for Good~0 under the CES DGP. \textit{Left}:
  Neural Demand (static). \textit{Right}: LA-AIDS. The neural demand system achieves near-exact
  prediction across the full support; LA-AIDS exhibits a systematic fan pattern reflecting
  misspecification of the CES Marshallian demand curvature.}
  \label{fig:sim_obs_pred}
\end{figure}

Figure~\ref{fig:sim_obs_pred} and Table~\ref{tab:sim_dgp_ces} show that the neural demand system
recovers the CES share mapping with near-exact accuracy, including under the post-shock
counterfactual. In the observed-versus-predicted plot, ND (static) lies essentially on the 45-degree
line over the full support, whereas LA-AIDS exhibits a fan pattern: it overpredicts at low observed
shares and underpredicts at high shares, consistent with curvature misspecification. Quantitatively,
the post-shock RMSE of ND (static) is $0.00063$ (s.e.\ $0.00004$), an order of magnitude smaller
than LA-AIDS and QUAIDS (both $\approx 0.0098$) and about five times smaller than the series
estimator ($0.0030$). The control-function variant performs similarly ($0.00075$), as expected in a
setting without endogeneity, while habit augmentation slightly degrades fit because the additional
state is redundant in a static DGP.

\begin{figure}[htbp]
  \centering
  \begin{subfigure}[b]{0.32\textwidth}
    \centering
    \includegraphics[width=\textwidth]{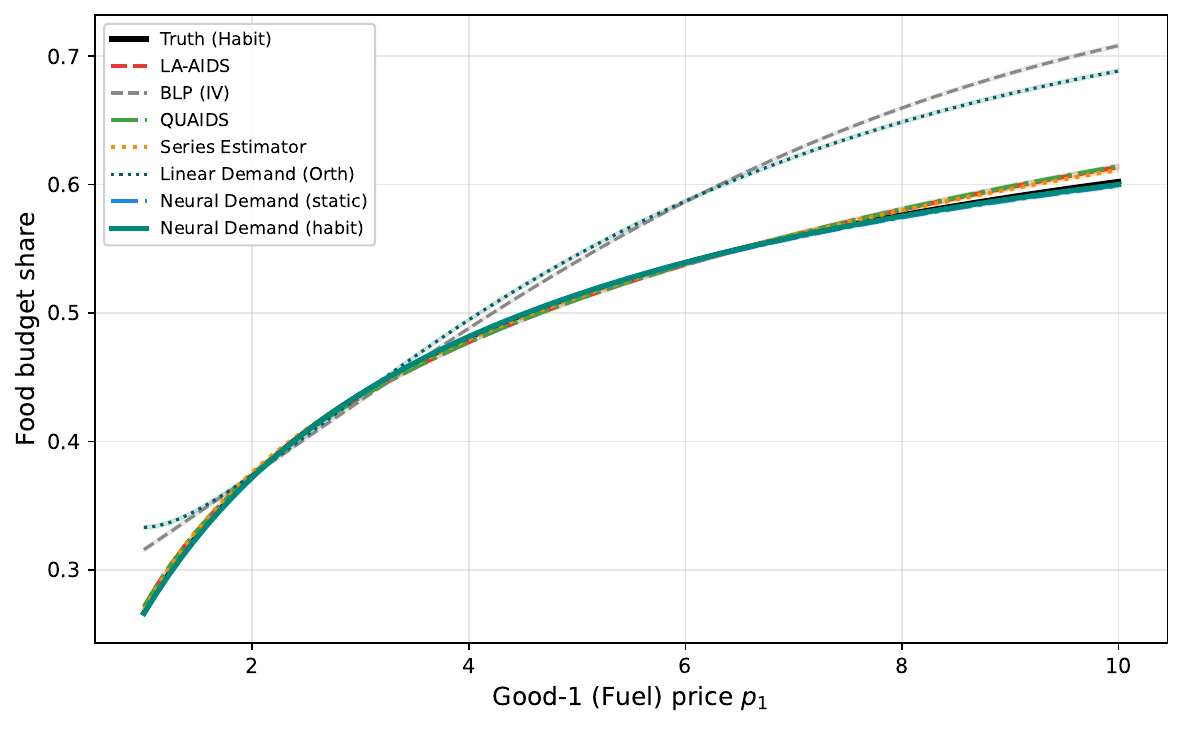}
    \caption{Good~0 (Food)}
    \label{fig:ces_good0}
  \end{subfigure}
  \hfill
  \begin{subfigure}[b]{0.32\textwidth}
    \centering
    \includegraphics[width=\textwidth]{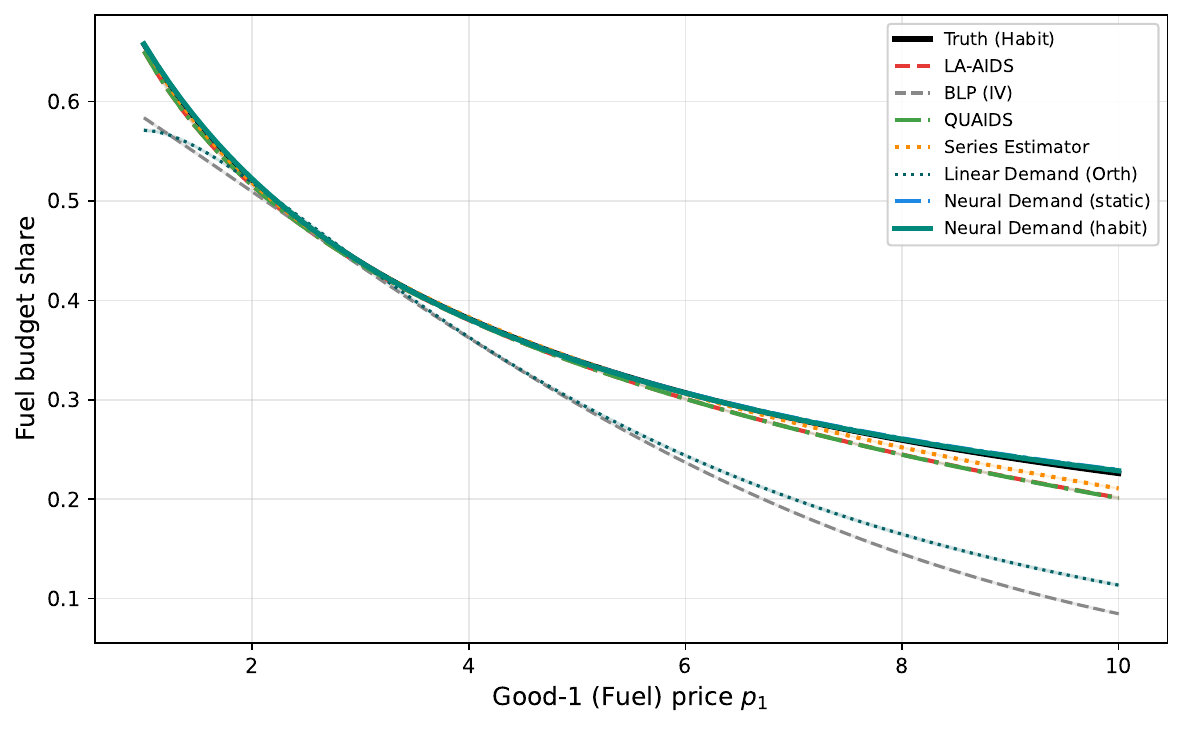}
    \caption{Good~1 (Fuel)}
    \label{fig:ces_good1}
  \end{subfigure}
  \hfill
  \begin{subfigure}[b]{0.32\textwidth}
    \centering
    \includegraphics[width=\textwidth]{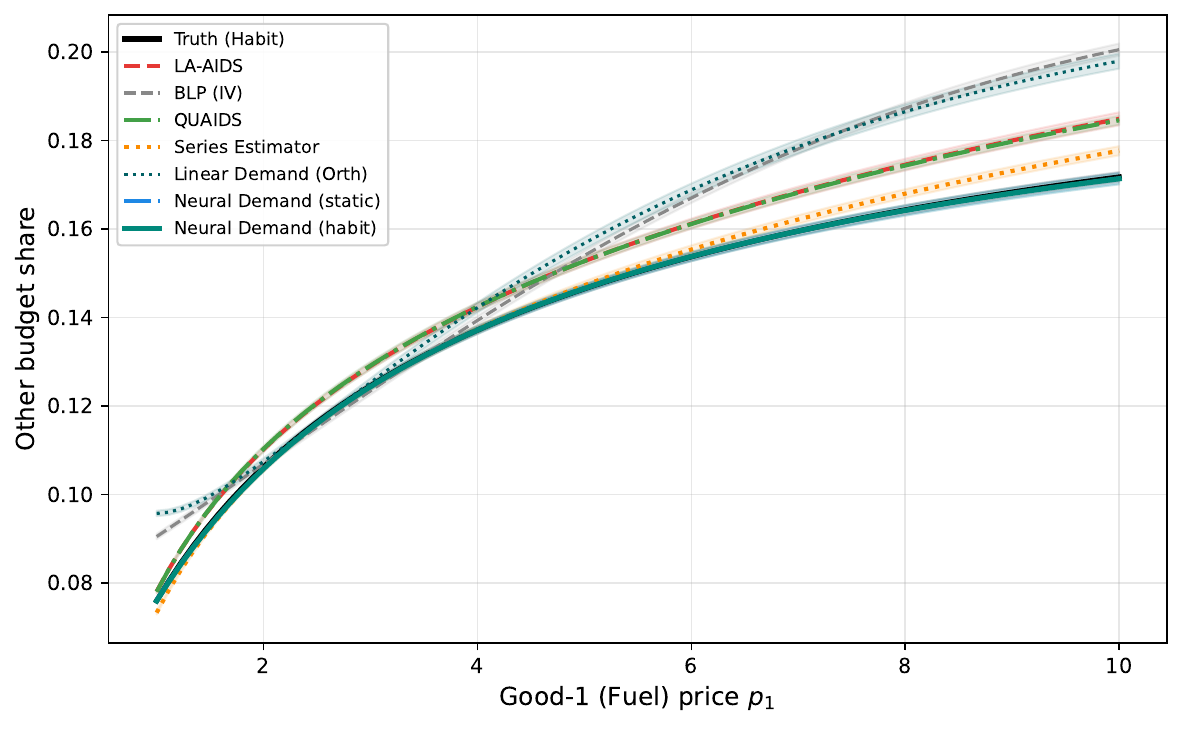}
    \caption{Good~2 (Other)}
    \label{fig:ces_good2}
  \end{subfigure}
  \caption{Predicted budget-share demand curves under the CES DGP ($\pm$1 SE across runs).
  Budget shares plotted as a function of Good-1 (Fuel) price $p_1$; all other prices and
  income held at simulation means.}
  \label{fig:sim_demand_ces}
\end{figure}

Figure~\ref{fig:sim_demand_ces} corroborates these findings by comparing implied demand curves as
$p_1$ varies with other prices and income fixed. ND (static) and ND (habit) closely track the CES
truth across all three goods and throughout the price range, indicating that the network captures not
only pointwise prediction but the correct global shape of substitution responses. Standard flexible
parametric benchmarks (LA-AIDS/QUAIDS and the series estimator) also approximate the truth well
over the interior of the support, but their larger post-shock RMSE indicates that small systematic
errors in curvature translate into noticeably larger counterfactual prediction errors. In contrast,
restrictive or misspecified structures fail: linear demand systems with shared or goodspecific
restrictions perform poorly, and discrete-choice comparators (BLP) diverge at the tails in this
continuous-allocation environment, reflecting their outside-option/IIA structure. Overall, the CES
experiment demonstrates that KL estimation on the simplex yields a high-fidelity approximation to
the structural share mapping and remains accurate under counterfactual price changes, providing a
clean benchmark for evaluating state-dependent extensions in later designs.

\paragraph{Elasticity Recovery.}

\begin{figure}[htbp]
  \centering
  \includegraphics[width=\linewidth]{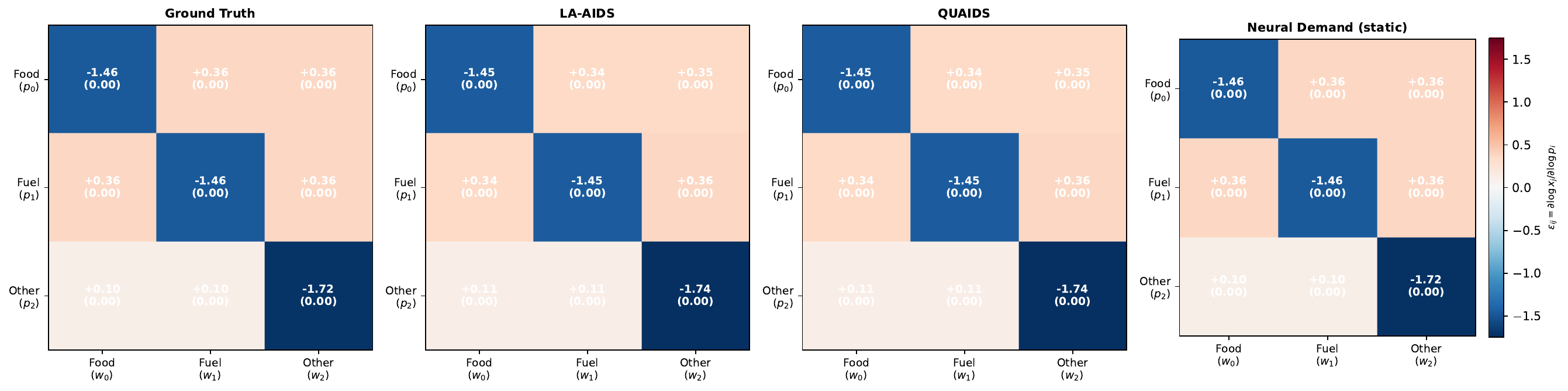}
  \caption{Full $3 \times 3$ price elasticity matrices under the CES DGP for Ground Truth,
  LA-AIDS, QUAIDS, and Neural Demand (static). Diagonal: own-price quantity elasticities.
  Off-diagonal: cross-price quantity elasticities. Neural Demand recovers both diagonal and
  off-diagonal elements with near-exact accuracy; LA-AIDS and QUAIDS produce small biases on
  the Good-2 own-price elasticity and the Food-Other cross-price terms.}
  \label{fig:sim_elasticity_heatmaps}
\end{figure}

Figure~\ref{fig:sim_elasticity_heatmaps} and Table~\ref{tab:sim_elasticities} evaluate recovery of the
CES substitution pattern when ground truth is known. The heatmaps show the full $3\times 3$ price
elasticity matrix and therefore test not only own-price slopes (diagonal) but also cross-price
substitution (off-diagonals). The CES truth has symmetric positive cross effects between Food and
Fuel (about $+0.36$), smaller cross effects involving the ``Other'' good (about $+0.10$), and more
elastic own-price response for Other (about $-1.71$) than for Food and Fuel (about $-1.44$).

Neural Demand (static) recovers this structure essentially exactly. In Table~\ref{tab:sim_elasticities},
its own-price elasticities match the ground truth within sampling error on all three goods
(e.g., $-1.440$ vs.\ $-1.438$ for Good~0; $-1.490$ vs.\ $-1.489$ for Good~1; $-1.718$ vs.\ $-1.710$ for
Good~2), and the heatmap shows the correct pattern and magnitude of cross effects. Habit augmentation
does not improve performance in this static DGP and yields virtually identical elasticities, as expected.
Similarly, the control-function variants match the baseline neural model, consistent with the absence
of endogeneity in this design.

\begin{table}[htbp]
  \centering
  \caption{Own-Price Quantity Elasticities — Simulation (CES DGP)}
  \label{tab:sim_elasticities}
  \begin{threeparttable}
  \begin{tabular}{lccc}
    \toprule
    \textbf{Model} & $\hat{\varepsilon}_{00}$ & $\hat{\varepsilon}_{11}$ & $\hat{\varepsilon}_{22}$ \\
    \midrule
    LA-AIDS & $-1.422 \pm 0.001$ & $-1.487 \pm 0.001$ & $-1.697 \pm 0.003$ \\
    BLP (IV) & $-1.485 \pm 0.004$ & $-1.662 \pm 0.005$ & $-1.787 \pm 0.004$ \\
    QUAIDS & $-1.421 \pm 0.001$ & $-1.487 \pm 0.001$ & $-1.699 \pm 0.003$ \\
    Series Estm. & $-1.424 \pm 0.001$ & $-1.478 \pm 0.001$ & $-1.713 \pm 0.001$ \\
    Linear Demand (Shared) & $-1.522 \pm 0.008$ & $-1.613 \pm 0.012$ & $-1.519 \pm 0.013$ \\
    Linear Demand (GoodSpec) & $-1.560 \pm 0.007$ & $-1.703 \pm 0.011$ & $-1.555 \pm 0.014$ \\
    Linear Demand (Orth) & $-1.515 \pm 0.006$ & $-1.688 \pm 0.006$ & $-1.838 \pm 0.004$ \\
    Neural Demand (static) & $-1.440 \pm 0.002$ & $-1.490 \pm 0.001$ & $-1.718 \pm 0.001$ \\
    Neural Demand (habit) & $-1.441 \pm 0.000$ & $-1.493 \pm 0.001$ & $-1.716 \pm 0.001$ \\
    Neural Demand (CF) & $-1.441 \pm 0.001$ & $-1.491 \pm 0.001$ & $-1.719 \pm 0.002$ \\
    Neural Demand (habit, CF) & $-1.432 \pm 0.002$ & $-1.480 \pm 0.003$ & $-1.695 \pm 0.002$ \\
    Ground Truth & $-1.438 \pm 0.001$ & $-1.489 \pm 0.001$ & $-1.710 \pm 0.001$ \\
    \bottomrule
  \end{tabular}
  \end{threeparttable}
\end{table}

Classical flexible share systems (LA-AIDS and QUAIDS) also perform well on the CES benchmark,
recovering the main diagonal accurately, but exhibit small systematic deviations for the Other good and
for cross terms (visible in the heatmaps). The series estimator is close but slightly noisier in the tails.
In contrast, linear-demand restrictions can distort the CES structure: the shared and goodspecific linear
systems produce implausibly similar elasticities across goods and misallocate substitution toward the
restricted dimensions, while the orthogonalized linear system improves but still departs from the CES
pattern. Overall, the CES experiment demonstrates that the neural demand system can recover the full
Slutsky-relevant substitution structure---not only own-price elasticities but also cross-price terms---in a
setting where the economic truth is known.

\paragraph{Compensating Variation.}

Table~\ref{tab:sim_welfare} evaluates recovery of compensating variation (CV) for a 20\% increase in
the price of Good~1 under the CES DGP, where the true CV is available in closed form. The neural
demand system matches the welfare benchmark essentially exactly: ND (static) and ND (CF) deliver
CV estimates statistically indistinguishable from the ground truth, with effectively zero percentage
error. Habit augmentation remains accurate (0.1\% error), as expected in a static DGP where the habit
state is redundant.

\begin{table}[htbp]
  \centering
  \caption{Compensating Variation — Simulation (CES DGP, 20\% shock)}
  \label{tab:sim_welfare}
  \begin{threeparttable}
  \begin{tabular}{lcc}
    \toprule
    \textbf{Model} & \textbf{CV} & \textbf{Error vs Truth (\%)} \\
    \midrule
    LA-AIDS & $-122.0925 \pm 0.2934$ & 0.4\% \\
    BLP (IV) & $-120.4891 \pm 0.3883$ & 1.7\% \\
    QUAIDS & $-122.1117 \pm 0.3041$ & 0.4\% \\
    Series Estm. & $-122.8321 \pm 0.2975$ & 0.2\% \\
    Linear Demand (Shared) & $-92.5375 \pm 0.2241$ & 24.5\% \\
    Linear Demand (GoodSpec) & $-92.0899 \pm 0.2394$ & 24.9\% \\
    Linear Demand (Orth) & $-121.3341 \pm 0.4326$ & 1.1\% \\
    Neural Demand (static) & $-122.6129 \pm 0.2892$ & 0.0\% \\
    Neural Demand (habit) & $-122.5390 \pm 0.3189$ & 0.1\% \\
    Neural Demand (CF) & $-122.6109 \pm 0.2814$ & 0.0\% \\
    Neural Demand (habit, CF) & $-122.6333 \pm 0.3410$ & 0.5\% \\
    Ground Truth & $-122.6243 \pm 0.2916$ & --- \\
    \bottomrule
  \end{tabular}
  \end{threeparttable}
\end{table}

Classical flexible share systems (LA-AIDS and QUAIDS) also recover welfare well (about 0.4\% error),
and the series estimator improves slightly (0.2\% error). By contrast, restrictive linear-demand
specifications can generate large welfare distortions: the shared and goodspecific linear systems
understate the welfare loss by roughly one quarter (24.5--24.9\% error), consistent with their
mis-shaped demand responses. The orthogonalized linear system performs substantially better (1.1\%
error), indicating that much of the welfare failure of the restricted linear systems is driven by their
inability to flexibly allocate price sensitivity across goods. Overall, the table shows that accurate
counterfactual welfare measurement requires a demand system that matches the shape of substitution
responses, and that KL estimation on the simplex can deliver welfare accuracy comparable to (and in
this benchmark, slightly better than) standard parametric share systems.

\subsubsection{Robustness Across Data-Generating Processes}
\label{sec:sim_robustness}

\begin{figure}[htbp]
  \centering
  \includegraphics[width=\linewidth]{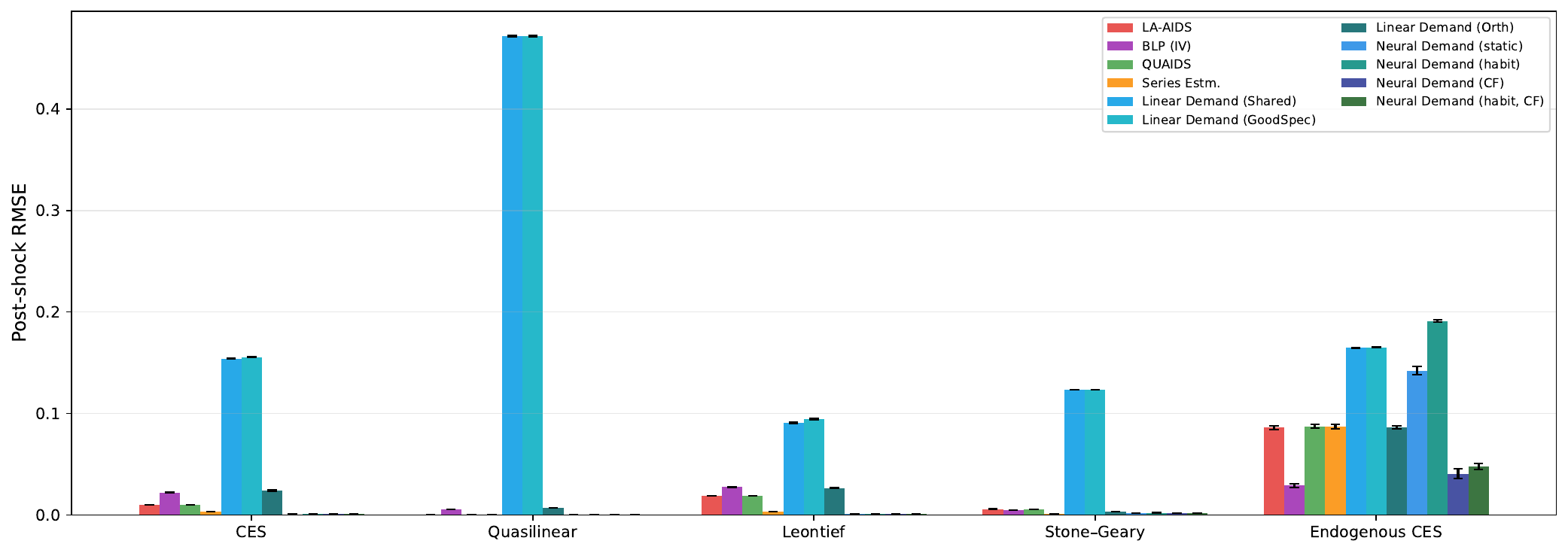}
  \caption{Post-shock RMSE across five DGPs ($\pm$SE, 5 runs). LDS (Shared) and LDS (GoodSpec)
  dominate the Quasilinear column at RMSE $\approx 0.47$; Var. Mixture dominates Leontief at
  $0.17$. Neural demand variants are essentially invisible on static DGPs. BLP (IV) is included
  for reference only; it is not an appropriate structural comparator for the continuous-allocation
  DGPs (Section~\ref{sec:blp}).}
  \label{fig:sim_dgp_robustness}
\end{figure}

\begin{figure}[htbp]
  \centering
  \includegraphics[width=\linewidth]{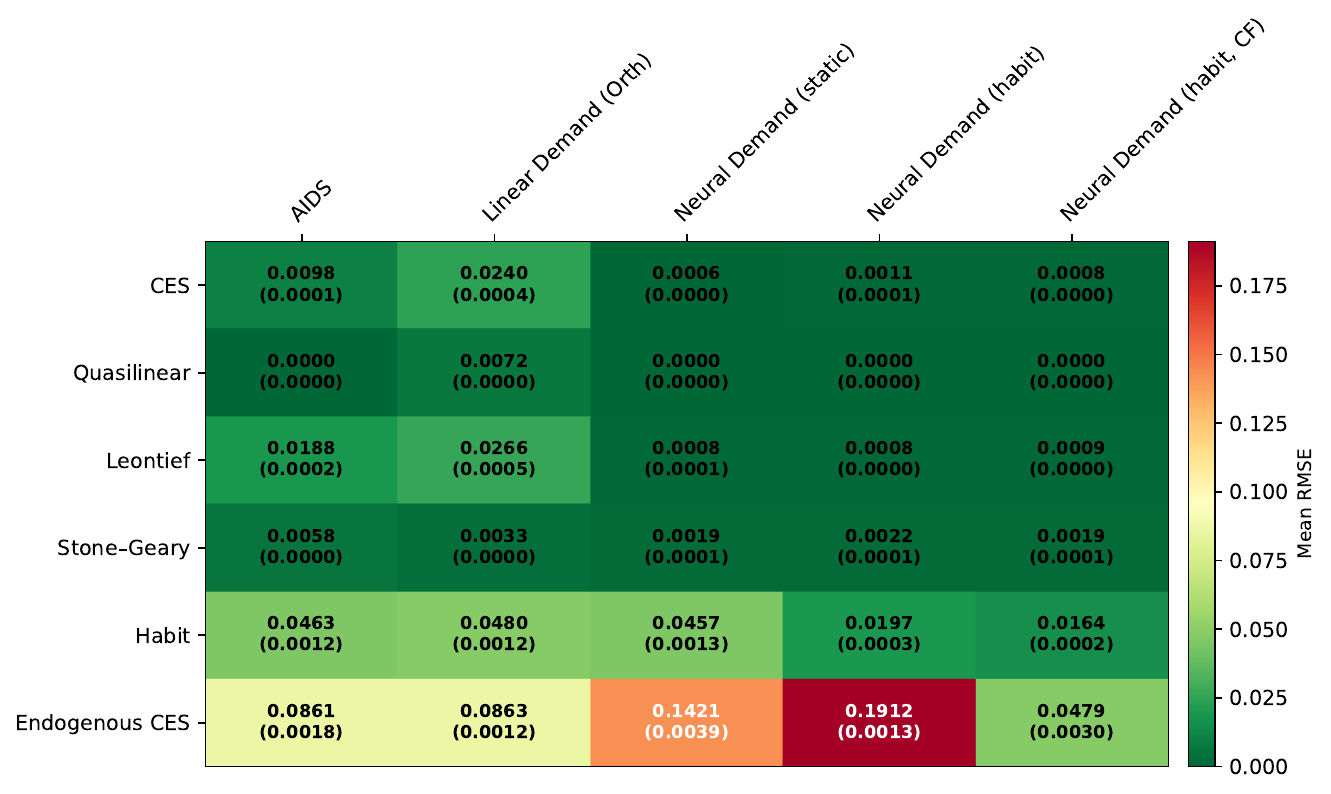}
  \caption{RMSE heatmap: key model comparators across six DGPs (mean RMSE, SE in parentheses).
  Red = high error; green = low error.}
  \label{fig:sim_rmse_heatmap}
\end{figure}

Table~\ref{tab:sim_dgp_others} and Figures~\ref{fig:sim_dgp_robustness} and \ref{fig:sim_rmse_heatmap} report post-shock RMSE
across four of the remaining DGPs: Quasilinear, Leontief, Stone-Geary and Endogenous CES. Four conclusions follow.

\begin{table}[htbp]
  \centering
  \caption{Post-Shock RMSE: Other DGPs (mean $\pm$ SE, 5 runs)}
  \label{tab:sim_dgp_others}
  \begin{threeparttable}
  \begin{tabular}{lcccc}
    \toprule
    \textbf{Model} & \textbf{Quasilinear} & \textbf{Leontief} & \textbf{Stone–Geary} & \textbf{Endog. CES} \\
    \midrule
    LA-AIDS & $0.00003 \pm 0.00000$ & $0.01884 \pm 0.00019$ & $0.00581 \pm 0.00004$ & $0.08610 \pm 0.00178$ \\
    BLP (IV) & $0.00577 \pm 0.00000$ & $0.02762 \pm 0.00030$ & $0.00478 \pm 0.00003$ & $0.02902 \pm 0.00189$ \\
    QUAIDS & $0.00003 \pm 0.00000$ & $0.01879 \pm 0.00022$ & $0.00576 \pm 0.00005$ & $0.08721 \pm 0.00182$ \\
    Series Estm. & $0.00003 \pm 0.00000$ & $0.00319 \pm 0.00002$ & $0.00112 \pm 0.00001$ & $0.08696 \pm 0.00248$ \\
    LD (Shared) & $0.47159 \pm 0.00079$ & $0.09059 \pm 0.00052$ & $0.12343 \pm 0.00012$ & $0.16474 \pm 0.00038$ \\
    LD (GoodSpec) & $0.47185 \pm 0.00076$ & $0.09465 \pm 0.00060$ & $0.12352 \pm 0.00013$ & $0.16524 \pm 0.00043$ \\
    LD (Orth) & $0.00718 \pm 0.00002$ & $0.02658 \pm 0.00046$ & $0.00330 \pm 0.00003$ & $0.08629 \pm 0.00120$ \\
    ND (static) & $0.00001 \pm 0.00000$ & $0.00076 \pm 0.00008$ & $0.00188 \pm 0.00009$ & $0.14214 \pm 0.00390$ \\
    ND (habit) & $0.00001 \pm 0.00000$ & $0.00076 \pm 0.00002$ & $0.00215 \pm 0.00006$ & $0.19115 \pm 0.00127$ \\
    ND (CF) & $0.00001 \pm 0.00000$ & $0.00082 \pm 0.00005$ & $0.00194 \pm 0.00012$ & $0.04062 \pm 0.00472$ \\
    ND (habit, CF) & $0.00001 \pm 0.00000$ & $0.00086 \pm 0.00005$ & $0.00186 \pm 0.00005$ & $0.04794 \pm 0.00299$ \\
    \bottomrule
  \end{tabular}
  \begin{tablenotes}\small
    \item Post-shock: 20\% increase in price of good 1. LD: Linear Demand; ND: Neural Demand.
  \end{tablenotes}
  \end{threeparttable}
\end{table}

\textit{Neural demand dominates on static nonlinear DGPs.} Across the CES, Leontief, and Stone--Geary
rows, Neural Demand (static) attains the lowest (or essentially lowest) post-shock RMSE among the
structural demand estimators, with especially large gains in environments with sharp nonlinearities.
In particular, it reduces RMSE from about $0.019$ (LA-AIDS/QUAIDS) to about $0.0008$ in Leontief
and from about $0.0098$ to about $0.0006$ in CES, consistent with Figure~\ref{fig:sim_dgp_robustness}
(top panel), where the neural bars are visually near-zero relative to classical share systems. The RMSE
heatmap in Figure~\ref{fig:sim_rmse_heatmap} reinforces this pattern: the CES and Leontief rows turn
uniformly dark green only for the neural specifications, indicating robust counterfactual prediction
under price shocks.

\textit{Static benchmarks perform well only when the mapping is easy.} In the quasilinear DGP,
multiple flexible estimators achieve near-zero RMSE because the true share mapping is close to linear
in the relevant region. The main failures in this row are the restricted linear-demand specifications:
LD (Shared) and LD (GoodSpec) exhibit catastrophic error (RMSE $\approx 0.47$), which dominates the
bar chart and appears as the brightest cells in the heatmap. This failure is structural: the imposed
restrictions prevent the model from accommodating the quasilinear environment’s negligible income
effects.

\textit{State dependence, not functional form, drives the Habit DGP.} The Habit row isolates the core
economic message of the paper: when the data-generating process is state dependent, purely static
models—no matter how flexible—cannot recover the post-shock mapping. LA-AIDS, Linear Demand,
and Neural Demand (static) all cluster at high RMSE (about $0.046$), shown by the uniformly red
Habit row in Figure~\ref{fig:sim_rmse_heatmap}. Augmenting the state with the habit stock produces a
discrete improvement: Neural Demand (habit) reduces RMSE to about $0.020$, and Neural Demand
(habit, CF) further to about $0.016$. This gap between identical neural architectures that differ only
in whether the habit state is included isolates the structural value of state augmentation.

\textit{Endogeneity correction is essential when prices are endogenous.} The Endogenous CES row
shows that endogeneity can dominate functional-form considerations. The uncorrected neural model
performs poorly (RMSE $0.142$) and the habit-augmented model is worse ($0.191$), reflecting that
adding state cannot resolve simultaneity between price and unobserved demand shocks. The
control-function correction restores predictive validity: Neural Demand (CF) reduces RMSE to $0.041$,
and Neural Demand (habit, CF) to $0.048$. This improvement is visible in both figures: the CF bars are
dramatically smaller in the bar chart and the Endogenous CES row turns from orange/red to green only
under the CF specifications. Overall, Figures~\ref{fig:sim_dgp_robustness}--\ref{fig:sim_rmse_heatmap}
show that credible counterfactual prediction requires matching the economic structure generating the
residual variation: state dependence requires a state variable, and endogeneity requires an explicit
correction.

\subsubsection{State Augmentation: Habit Formation Results}
\label{sec:sim_habit}

\paragraph{Demand Curves.}

\begin{figure}[htbp]
  \centering
  \begin{subfigure}[b]{0.32\textwidth}
    \centering
    \includegraphics[width=\textwidth]{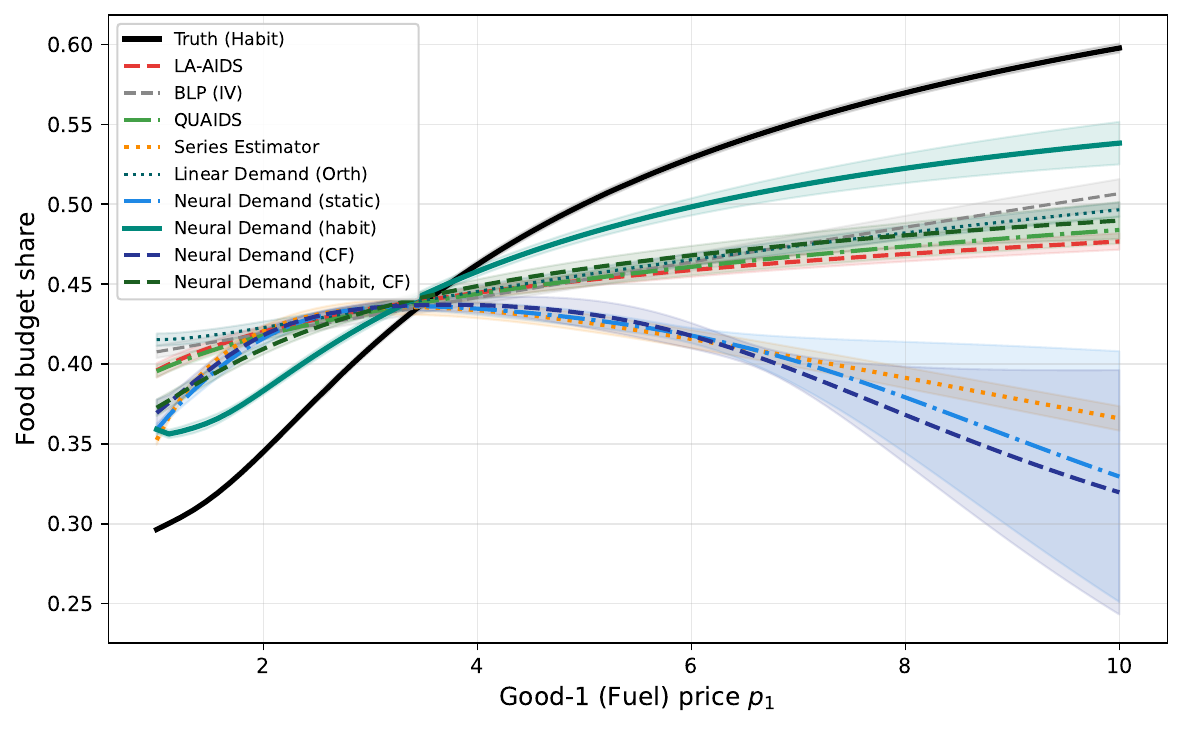}
    \caption{Good~0 (Food)}
    \label{fig:habit_good0}
  \end{subfigure}
  \hfill
  \begin{subfigure}[b]{0.32\textwidth}
    \centering
    \includegraphics[width=\textwidth]{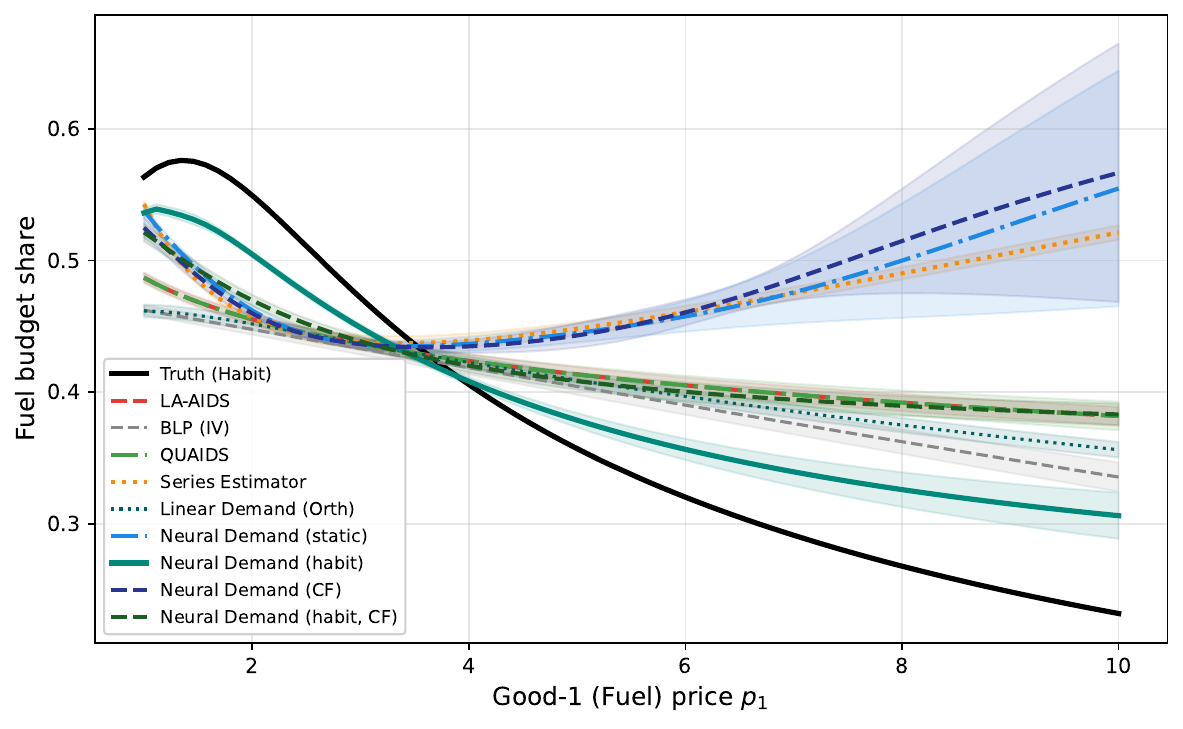}
    \caption{Good~1 (Fuel)}
    \label{fig:habit_good1}
  \end{subfigure}
  \hfill
  \begin{subfigure}[b]{0.32\textwidth}
    \centering
    \includegraphics[width=\textwidth]{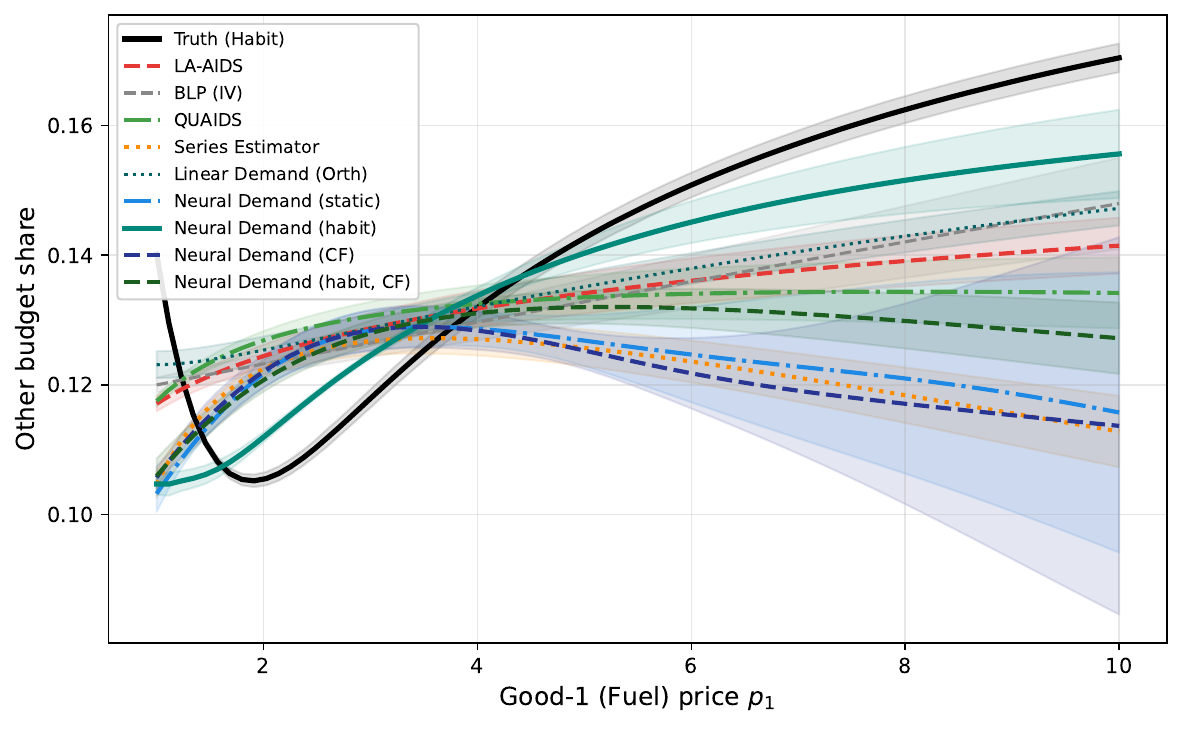}
    \caption{Good~2 (Other)}
    \label{fig:habit_good2}
  \end{subfigure}
  \caption{Predicted budget-share demand curves under the Habit Formation DGP ($\theta=0.3$,
  $\delta_{\text{DGP}}=0.7$; habit stocks held at training means). The true Fuel demand (Good~1,
  centre) exhibits a non-monotone shape driven by habit persistence that all static models miss.
  Neural Demand (habit) qualitatively captures the pattern. BLP is included for reference only;
  it is not an appropriate structural comparator for this DGP.}
  \label{fig:sim_demand_habit}
\end{figure}

Figure~\ref{fig:sim_demand_habit} plots budget-share demand curves as the Fuel price $p_1$ varies,
holding other prices, income, and the habit stock at their training means. The habit DGP generates
qualitatively different comparative statics than any static model can represent: in the Fuel panel
(Good~1), the truth exhibits a pronounced non-monotonic response (an interior peak at low $p_1$,
followed by a steep decline), while static estimators---including flexible share systems and Neural
Demand (static)---produce smooth, monotone curves that miss this shape. Neural Demand (habit)
moves substantially toward the truth across all three goods, capturing the steepening of the Fuel
response and the associated reallocation into Food and Other at higher $p_1$. The remaining gap at
very low $p_1$ is consistent with partial identification of the decay parameter and with the fact that
the curve is traced at a fixed habit stock rather than along the endogenous state distribution.

\paragraph{Habit-State Augmentation Advantage.}

\begin{figure}[htbp]
  \centering
  \includegraphics[width=\linewidth]{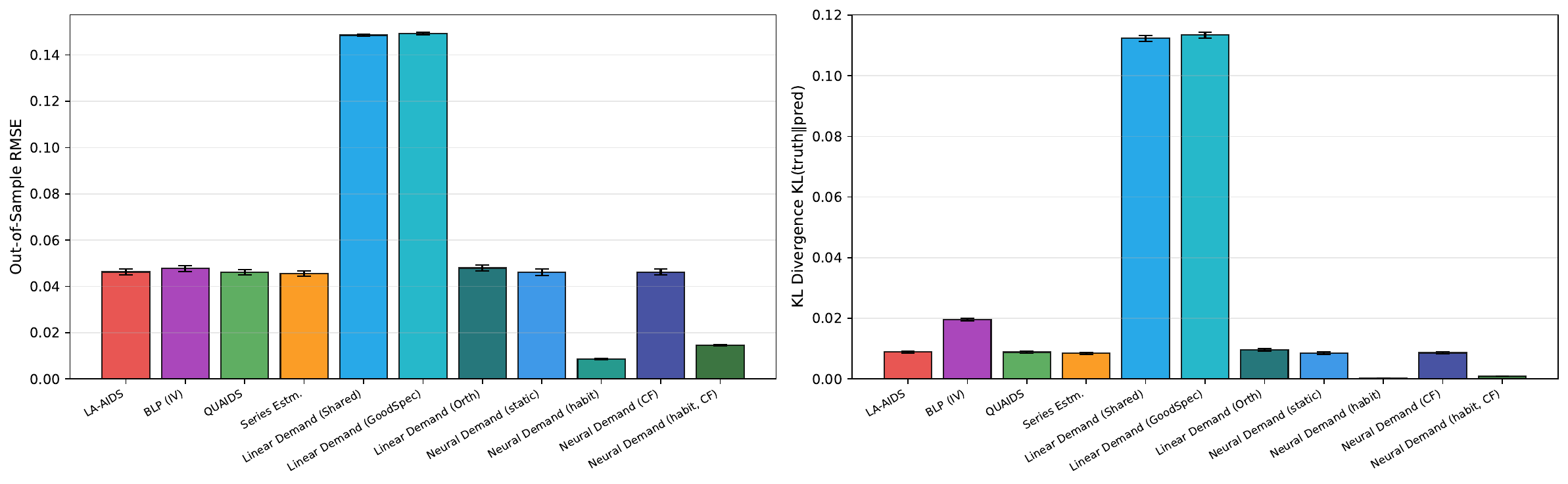}
  \caption{Habit-state augmentation advantage on the Habit Formation DGP. \textit{Left}: RMSE.
  \textit{Right}: KL divergence. All static models cluster at RMSE $\approx 0.047$--$0.050$
  (equivalent to LA-AIDS). Neural Demand (habit) reduces RMSE by 80\% and KL by a factor of
  more than 20.}
  \label{fig:sim_habit_advantage}
\end{figure}

Table~\ref{tab:sim_habit_advantage} and Figure~\ref{fig:sim_habit_advantage} quantify the
implications for counterfactual accuracy. All static specifications cluster near RMSE $0.046$ and KL
$\approx 0.009$, including Neural Demand (static) and Neural Demand (CF), showing that functional-form
flexibility and endogeneity correction do not address omitted state dependence. In contrast, adding the
habit state reduces RMSE by more than 80\% (to $0.0085$) and lowers KL by an order of magnitude
(to $3.3\times 10^{-4}$). The profile procedure selects $\hat\delta=0.653\pm 0.017$, and the true
$\delta_{\text{true}}=0.7$ lies in the identified set in all seeds, consistent with weak but informative
identification of the persistence horizon. Overall, the habit experiment isolates the paper's main
mechanism: when the DGP is state dependent, the first-order gains come from augmenting the state
with a sufficient statistic for history, not from additional static flexibility.

\begin{table}[htbp]
  \centering
  \caption{Habit-State Augmentation Advantage --- Simulation  (5 runs; mean $\pm$ SE)}
  \label{tab:sim_habit_advantage}
  \begin{threeparttable}
  \begin{tabular}{lccc}
    \toprule
    \textbf{Model} & \textbf{RMSE} & \textbf{KL Div.}    & \textbf{RMSE Red.\ (\%)}  \\
    \midrule
    LA-AIDS & $0.04625 \pm 0.00116$ & $0.00887 \pm 0.00039$ & 0.0\% \\
    BLP (IV) & $0.04771 \pm 0.00118$ & $0.01959 \pm 0.00041$ & -3.2\% \\
    QUAIDS & $0.04619 \pm 0.00117$ & $0.00885 \pm 0.00040$ & 0.1\% \\
    Series Estm. & $0.04553 \pm 0.00113$ & $0.00843 \pm 0.00037$ & 1.6\% \\
    Linear Demand (Shared) & $0.14856 \pm 0.00048$ & $0.11231 \pm 0.00094$ & -222.0\% \\
    Linear Demand (GoodSpec) & $0.14935 \pm 0.00052$ & $0.11339 \pm 0.00099$ & -223.7\% \\
    Linear Demand (Orth) & $0.04796 \pm 0.00119$ & $0.00953 \pm 0.00042$ & -3.7\% \\
    Neural Demand (static) & $0.04617 \pm 0.00135$ & $0.00861 \pm 0.00044$ & 0.2\% \\
    Neural Demand (habit) & $0.00852 \pm 0.00026$ & $0.00033 \pm 0.00002$ & 81.5\% \\
    Neural Demand (CF) & $0.04616 \pm 0.00126$ & $0.00863 \pm 0.00042$ & 0.2\% \\
    Neural Demand (habit, CF) & $0.01454 \pm 0.00018$ & $0.00088 \pm 0.00002$ & 68.5\% \\
    \bottomrule
  \end{tabular}
  \begin{tablenotes}\small
    \item True DGP: \texttt{HabitFormationConsumer} with    $\delta_{true}=0.7$.    $\hat{\delta} = 0.653 \pm    0.017$ (mean $\pm$ SE over 5 runs).    True $\delta$ in identified set: 100\% of seeds.
  \end{tablenotes}
  \end{threeparttable}
\end{table}

The habit-formation simulation isolates the paper’s central mechanism. When the true demand
process is state dependent, static demand estimators---including highly flexible nonparametric share
systems---cannot recover counterfactual responses to price shocks, because the dominant source of
variation is omitted history rather than static curvature. The demand-curve comparisons show that
static models miss the characteristic non-monotonic and steepened responses induced by habits,
whereas the habit-augmented neural model moves substantially toward the truth across goods. This
translates into large gains in counterfactual accuracy: adding the habit state reduces post-shock RMSE
by more than 80\% and lowers KL divergence by an order of magnitude relative to static benchmarks.
Finally, profiling over the decay parameter yields a long-memory regime and an identified set that
contains the true $\delta$ in all seeds, indicating that while the persistence horizon is only weakly
pinned down, it is sufficiently identified to recover elasticities and welfare objects accurately in this
environment.

\paragraph{Habit-Decay Parameter Recovery.}
\label{sec:sim_delta}

\begin{figure}[htbp]
  \centering
  \includegraphics[width=0.75\linewidth]{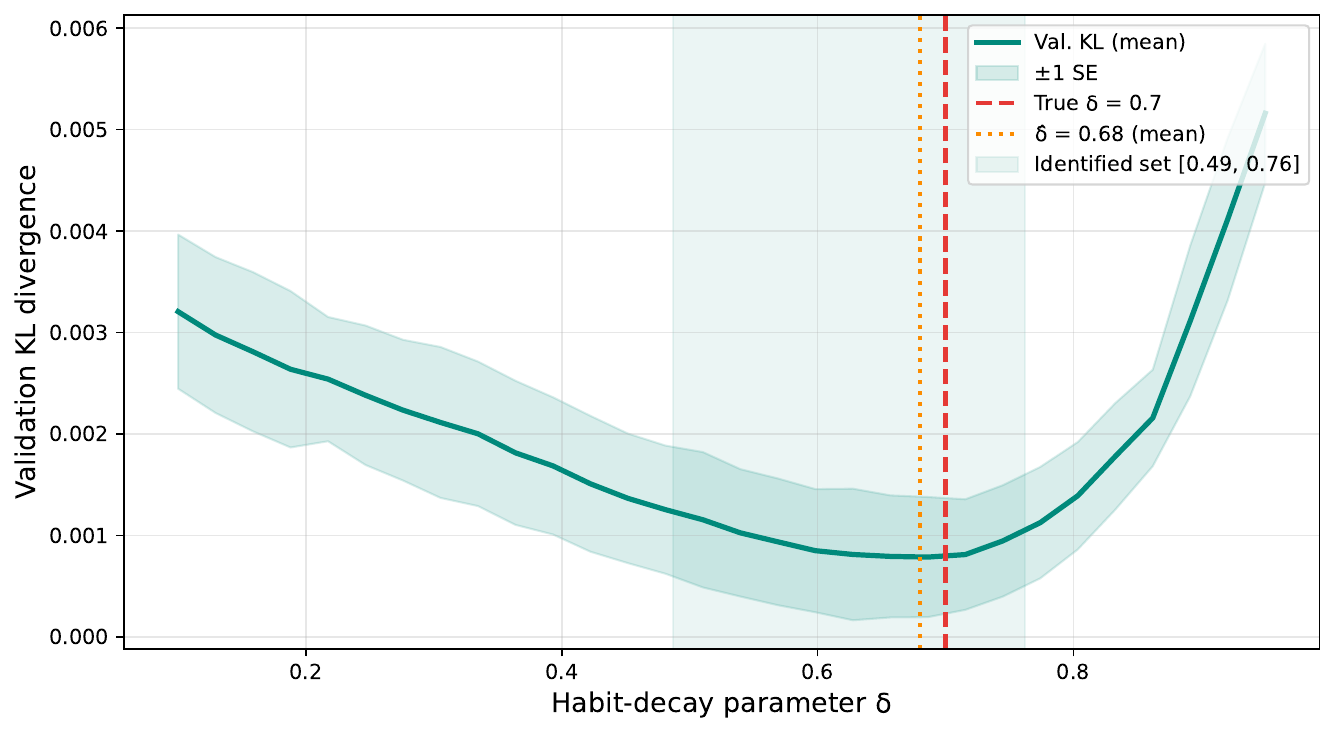}
  \caption{Profile KL criterion for $\delta$ identification ($\pm$1 SE, 5 seeds). The minimum
  coincides with the true value $\delta_{\text{DGP}} = 0.7$ (red dashed line). The flat region
  spanning approximately $[0.54, 0.77]$ under the 2-SE rule constitutes the identified set: values
  below 0.5 and above 0.85 are sharply rejected; intermediate values are observationally
  near-equivalent.}
  \label{fig:sim_habit_profile_kl}
\end{figure}

Figure~\ref{fig:sim_habit_profile_kl} and Table~\ref{tab:sim_delta_identification} summarize identification
of the habit-decay parameter $\delta$ in the habit-formation simulation using the profile (validation)
KL criterion. The profile curve is U-shaped with a clear interior minimum near the true value:
the mean minimizer is $\hat\delta=0.680$ (s.e.\ $0.014$) against $\delta_{\text{true}}=0.7$. At the same
time, the criterion is relatively flat around the minimum, yielding a nontrivial identified set.
Using the threshold rule $\widehat{KL}_{\min}+2\,\widehat{SE}$, the average identified set is
$[0.487,\,0.762]$ with mean width $0.276$ (s.e.\ $0.122$), reflecting weak (set) identification of the
precise persistence horizon.

Two conclusions follow. First, the profiling procedure is informative about the \emph{regime} of
persistence: low values of $\delta$ are rejected, and the minimizer lies close to the true long-memory
value. Second, inference should be stated as set-valued: a range of decay rates deliver
observationally near-equivalent fit because the resulting EWMA habit stocks are highly collinear on
the observed support. Consistent with this interpretation, the identified set contains the true
$\delta=0.7$ in 100\% of seeds. Practically, this motivates reporting substantive objects (elasticities,
decompositions, welfare) either at $\hat\delta$ and/or as ranges over the identified set, rather than as
point estimates that overstate precision about the exact memory length.

\begin{table}[htbp]
  \centering
  \caption{Partial Identification of $\delta$ (Habit-Decay Parameter) --- 5 seeds}
  \label{tab:sim_delta_identification}
  \begin{tabular}{lc}
    \toprule
    \textbf{Quantity} & \textbf{Value} \\
    \midrule
    True $\delta$ & 0.7 \\
    $\hat{\delta}$ (mean ± SE) & 0.680 ± 0.014 \\
    IS lo (mean) & 0.487 \\
    IS hi (mean) & 0.762 \\
    IS width (mean±SE) & 0.276 ± 0.122 \\
    Coverage (\%) & 100 \\
    \bottomrule
  \end{tabular}
  \caption*{Profile KL threshold = $\hat{KL}_{min} + 2\,\widehat{SE}$.  True $\delta=0.7$.  Identified set covers the truth in  100\% of simulation seeds.}
\end{table}

Profiling over the habit-decay parameter using validation KL yields a clear but set-valued conclusion.
The profile criterion has an interior minimum near the truth ($\hat\delta=0.680$ versus
$\delta_{\text{true}}=0.7$), rejecting short-memory values, but remains relatively flat in the
neighborhood of the minimum, producing an identified set of nontrivial width. Across seeds, the
average identified set $[0.487,0.762]$ covers the true decay rate in 100\% of runs, indicating that the
procedure reliably locates the correct persistence regime while acknowledging that the precise memory
length is only weakly pinned down. Accordingly, subsequent elasticities and welfare objects are most
appropriately reported at $\hat\delta$ and/or over the identified set.

\section{Empirical Application: Dominick's Analgesics}
\label{sec:data}

\subsection{Data and Construction}

We apply the neural demand system to weekly scanner data from Dominick's Finer Foods, covering 93 stores over 399 weeks and 641 UPCs in the OTC analgesics category (7,339,217 raw observations). Analgesics are a canonical repeat-purchase category in which persistence can arise from loyalty/habits, inventory dynamics (stockpiling during promotions), and stable heterogeneity across store catchment areas. The category also features substantial week-to-week price variation driven by promotions, making it a useful setting to evaluate whether incorporating a low-dimensional state changes implied substitution and welfare. We aggregate UPCs to three ``inside goods'' by active ingredient to focus on the economic distinction between analgesic types rather than on fine product differentiation.

\subsubsection{Product Aggregation.}
Products are classified into three active-ingredient groups:
\begin{enumerate}
    \item[(i)] Aspirin: Bayer, Bufferin, Anacin, Ascriptin, Ecotrin, BC Powder, and generics.
    \item[(ii)] Acetaminophen: Tylenol, Excedrin (aspirin-free), Panadol, Datril, Pamprin, Midol (non-ibuprofen), store brands.
    \item[(iii)] Ibuprofen: Advil, Motrin IB, Nuprin, Aleve, Actron, Orudis KT, and store brands.
\end{enumerate}
Unclassified products are excluded. The three-good aggregation preserves the primary economically meaningful substitution dimension --- between analgesic active ingredients --- while maintaining sufficient within-category revenue coverage for reliable share construction.

\subsubsection{Unit Prices.}
Following \citet{deaton1988quality}, prices are standardised to a \$100-tablet equivalent:
\[
\hat{p}_{ijt} = \text{PRICE}_{ijt} \times \frac{100}{\text{TABLETS}_j}.
\]
Prices are aggregated as revenue-weighted means within each store $\times$ week $\times$ category cell. This standardisation ensures that price comparisons across UPCs with different package sizes reflect per-unit cost rather than package-size variation.

\subsubsection{Budget Shares.}
Budget shares are given by 
\[
w_{igt} = \frac{R_{igt}}{\sum_g R_{igt}},
\]
where $R_{igt}$ is total revenue in store $i$, good $g$, week $t$.

\subsubsection{Income Proxy.}
Income is proxied by total within-category revenue: $y_{it} = \max\!\left(\sum_g R_{igt},\, 1\right) / 100$. This absorbs variation in store traffic but does not correspond to full household income. We note that this proxy is endogenous to the price shock used in welfare calculations; compensating variation is therefore computed holding the income proxy fixed at its pre-shock value. As discussed in Section~\ref{sec:identification}, welfare levels computed with this proxy are in units of within-category expenditure and are not interpretable as dollar household welfare losses; cross-model comparisons are ordinal.

\subsubsection{Habit Stock Construction.}
The habit stock is constructed in log-share space rather than quantity space, as budget shares are unit-invariant and comparable across stores. For each $\delta_k$ on the profile grid:
\[
  \log \bar{x}_t(\delta_k) = \delta_k \cdot \log \bar{x}_{t-1}(\delta_k) + (1 - \delta_k) \cdot \log w_{t-1},
\]
initialised at the cross-store global mean log-share and reset to that mean at store boundaries. The full sequence $\bar{x}(\delta_k)$ is pre-computed once per grid point before any network training. The profile criterion selects $\hat\delta$, and $\bar{x}(\hat\delta)$ is passed as a fixed input to the habit-augmented model.

\subsubsection{Train/Test Split.}
The balanced panel contains 32,931 store--week observations. Weeks $\geq 351$ are held out for testing (4,287 observations); weeks 1--350 are used for training (28,644 observations). This temporal split ensures that test-set predictions are genuinely out-of-sample: no information from weeks 351--399 influences model parameters.

\begin{table}[htbp]
  \centering
  \caption{Descriptive Statistics: Dominick's Analgesics Scanner Panel}
  \label{tab:dom_desc}
    \begin{tabular}{lcccc}
      \hline
      Good & Mean Price & Std Price & Mean Share & Std Share \\
      & {(\$/100 tablets)} & & & \\
      \hline
      Aspirin & 7.311 & 1.557 & 0.2107 & 0.0519 \\
      Acetaminophen & 10.554 & 1.427 & 0.5026 & 0.0785 \\
      Ibuprofen & 10.506 & 1.428 & 0.2867 & 0.0819 \\
      \hline
      \multicolumn{5}{l}{\textit{Train: 28,644 obs\quad Test: 4,287 obs\quad Stores: 93\quad Weeks: 1--399}} \\
      \hline
    \end{tabular}
    \caption*{Unit prices per-100-tablet equivalent. Revenue-weighted means within store$\times$week$\times$category.
      Aspirin: Bayer, Bufferin, Anacin, Ascriptin, Ecotrin, generics.
      Acetaminophen: Tylenol, Excedrin, Anacin-3, Panadol, Datril, Pamprin, store brands.
      Ibuprofen: Advil, Motrin~IB, Nuprin, Aleve, Actron, Orudis~KT, Haltran, Medipren.}
\end{table}

Acetaminophen dominates category expenditure (50\%), followed by ibuprofen (29\%) and aspirin (21\%). Acetaminophen and ibuprofen command nearly identical per-unit prices on average (\$10.55 and \$10.51 per 100 tablets, respectively), while aspirin is markedly cheaper at \$7.31. Price variation is substantial --- standard deviations of approximately \$1.4--\$1.6 across all three goods --- reflecting primarily store-level promotional activity rather than cross-store cost differences. This promotional variation is the source of the endogeneity concern addressed by the control function specification.

\subsubsection{Store Fixed Effects.}
We estimate store fixed-effects variants of the static and habit-augmented models, assigning each store a learnable 8-dimensional embedding concatenated to the input before the hidden layers. Store embeddings are initialised with $\sigma = 0.01$ to start near the no-fixed-effect baseline. These variants absorb time-invariant store-level heterogeneity --- including neighbourhood income, local brand preferences, and store format --- that could otherwise confound identification of the habit effect.

\subsubsection{Placebo.}

To verify that the performance gains from the habit/state-augmented specification reflect genuine
temporal dependence rather than a mechanical benefit from adding extra inputs, we implement a
placebo ``shuffled history'' test. The placebo model uses the same habit-augmented architecture and
training protocol as the baseline habit model, but it breaks the alignment between each observation
and its own consumption history by randomly permuting the history inputs across observations. In
particular, the habit-stock and lagged-quantity features $(\bar x_{i,t-1}(\hat\delta), q_{i,t-1})$ are
permuted across the training set prior to estimation, and independently permuted across the test set
prior to evaluation, while contemporaneous covariates $(p_{it},y_{it})$ are left unchanged. This
preserves the marginal distribution of the history features but eliminates their structural interpretation
as the state associated with $(i,t)$. If the habit model's gains were driven primarily by architecture or
generic overfitting to additional covariates, the placebo model would perform similarly to the true
habit model.

\subsection{Results}
\label{sec:empirical_results}

\subsubsection{Out-of-Sample Predictive Accuracy}
\label{sec:dom_accuracy}

\begin{figure}[htbp]
  \centering
  \includegraphics[width=\linewidth]{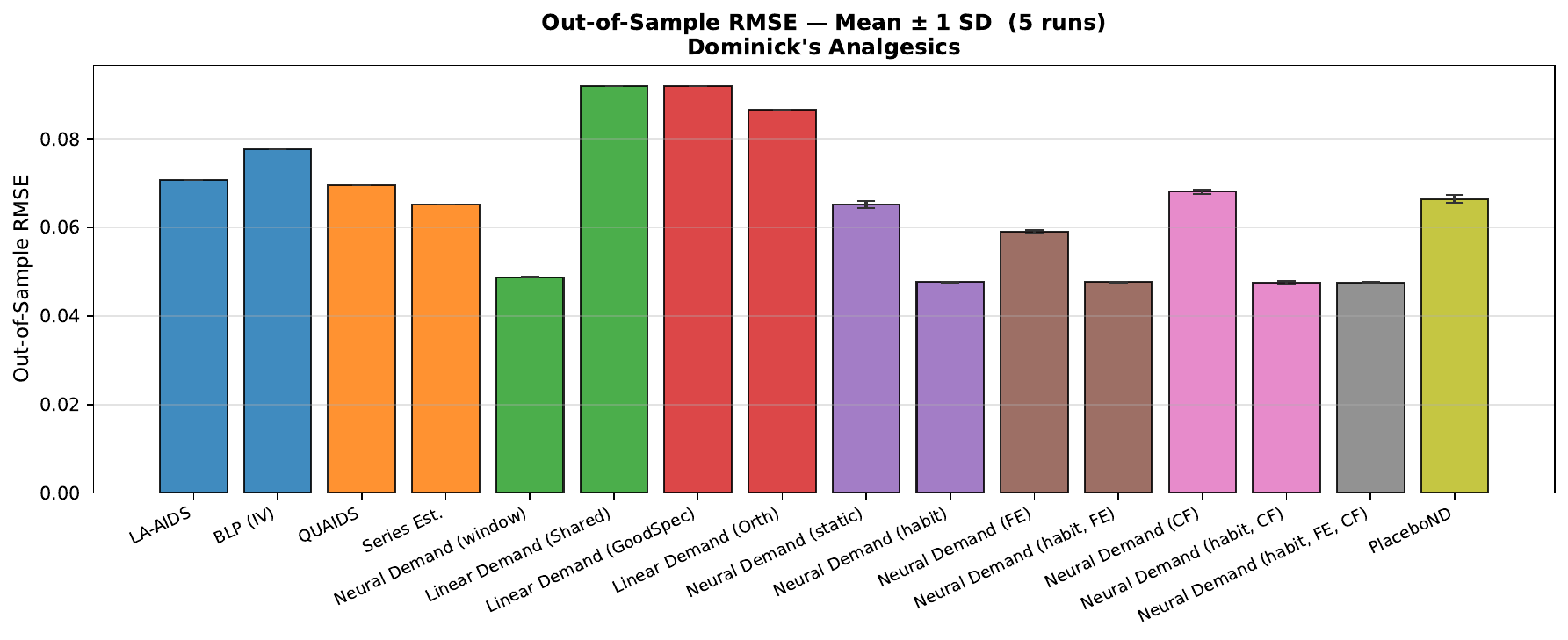}
  \caption{Out-of-sample RMSE --- Dominick's Analgesics. The habit-augmented models achieve RMSE $\approx 0.048$, approximately 33\% below the LA-AIDS baseline.}
  \label{fig:dom_accuracy}
\end{figure}

Table~\ref{tab:dom_acc} and Figure~\ref{fig:dom_accuracy} report out-of-sample RMSE and MAE on the held-out test weeks across all sixteen specifications. The results establish a clear three-tier ranking. The first-stage diagnostics for CF model are in the appendix (section \ref{sec:first_stage}). 

\begin{table}[htbp]
  \centering
  \caption{Out-of-Sample Predictive Accuracy --- Dominick's Analgesics (5 independent re-estimations; mean $\pm$ std)}
  \label{tab:dom_acc}
  \begin{threeparttable}
      \begin{tabular}{lcc}
       \toprule
    \textbf{Model} & \textbf{RMSE} & \textbf{MAE} \\
    \midrule
    LA-AIDS & $0.07069$ & $0.05593$  \\
    BLP (IV) & $0.07766$ & $0.05991$  \\
    QUAIDS & $0.06956$ & $0.05519 $  \\
    Series Est. & $0.06514$ & $0.05077$  \\
    Linear Demand (Shared) & $0.09201$ & $0.07361$  \\
    Linear Demand (GoodSpec) & $0.09197$ & $0.07356$  \\
    Linear Demand (Orth) & $0.08662$ & $0.07046$  \\
    Neural Demand (static) & $0.06521 \pm 0.00037$ & $0.05047 \pm 0.00031$  \\
    Neural Demand (window) & $0.04872 \pm 0.00004$ & $0.03661 \pm 0.00004$  \\
    Neural Demand (habit) & $0.04762 \pm 0.00007$ & $0.03574 \pm 0.00008$  \\
    Neural Demand (FE) & $0.05902 \pm 0.00021$ & $0.04506 \pm 0.00019$  \\
    Neural Demand (habit, FE) & $0.04763 \pm 0.00007$ & $0.03583 \pm 0.00005$  \\
    Neural Demand (CF) & $0.06806 \pm 0.00020$ & $0.05196 \pm 0.00010$  \\
    Neural Demand (habit, CF) & $0.04751 \pm 0.00015$ & $0.03572 \pm 0.00008$  \\
    Neural Demand (Placebo) & $0.06646 \pm 0.00091$ & $0.05139 \pm 0.00069$ \\
    \bottomrule
    \end{tabular}
    \begin{tablenotes}\small
      \item RMSE and MAE on held-out test observations, mean $\pm$ std over 5 runs. No std reported for the deterministic models. 
    \end{tablenotes}
  \end{threeparttable}
\end{table}

Table~\ref{tab:dom_acc} reports out-of-sample predictive accuracy on held-out Dominick's
store-weeks. Three messages are immediate. First, relative to classical continuous share systems,
the static neural demand model delivers a modest but consistent improvement: Neural Demand (static)
achieves RMSE $0.0652$ versus $0.0707$ for LA-AIDS and $0.0696$ for QUAIDS, and performs
comparably to the flexible series benchmark (RMSE $0.0651$). This indicates that in a purely static
specification, the gains from the neural parameterization are present but limited—consistent with the
fact that standard translog/spline bases already approximate local curvature reasonably well in this
category.

Second, incorporating state dependence yields a large first-order improvement in predictive accuracy.
The habit-augmented neural model reduces RMSE to approximately $0.0476$ (and MAE to $0.0357$),
a reduction of roughly one-third relative to LA-AIDS and about 27\% relative to the static neural
model. The magnitude of this improvement, together with its extremely small dispersion across
re-estimations, indicates that a substantial share of the residual variation in static models reflects
persistent dynamics that are well summarized by the constructed history state.

Third, the robustness patterns across variants help interpret the source of the gain. Adding store fixed
effects improves the static neural model (RMSE $0.0590$), confirming that time-invariant store
heterogeneity accounts for part of the unfit variation in static specifications, but fixed effects do not
materially alter the performance of the habit model (RMSE $0.0476$ with or without FE). Likewise,
the control-function correction slightly worsens predictive fit in the static model (RMSE $0.0681$),
consistent with its role as a causal correction rather than a forecasting device, yet the habit+CF
variant matches the baseline habit model (RMSE $0.0475$). Taken together, these results suggest
that the dominant empirical content of the state augmentation is not reducible to constant store
intercepts or to the endogenous component of price variation; rather, it reflects persistent
state dependence in demand that materially improves out-of-sample prediction.

Finally, the window model performs similarly to the habit specification (RMSE $0.0487$), indicating
that short-run lagged information is valuable in this category. The fact that the EWMA habit stock
slightly outperforms the richer window input is consistent with the view that a parsimonious
single-parameter summary can concentrate the relevant persistence while reducing the dimensionality
and noise of raw lagged inputs.

\paragraph{Placebo results and interpretation.}
The placebo specification is a falsification test for the interpretation that state augmentation captures
genuine temporal dependence rather than delivering gains mechanically from additional inputs. In the
placebo model, the history features (habit stock and lagged quantities) are randomly permuted across
observations in both training and testing, preserving their marginal distribution but destroying the
alignment between each store-week and its own consumption history. The placebo achieves RMSE
$0.06646\pm0.00091$ and MAE $0.05139\pm0.00069$, which is close to the static neural model’s error
and far above the habit model’s RMSE of about $0.0476$. Thus, simply providing history-shaped inputs
does not reproduce the habit model’s advantage; the improvement depends on the correct temporal
link between past consumption and current demand, consistent with economically meaningful
persistence (brand loyalty/state dependence) rather than a generic ``more covariates'' effect.

\subsubsection{Own-Price and Cross-Price Elasticities}
\label{sec:dom_elast}

\begin{figure}[htbp]
  \centering
  \includegraphics[width=\linewidth]{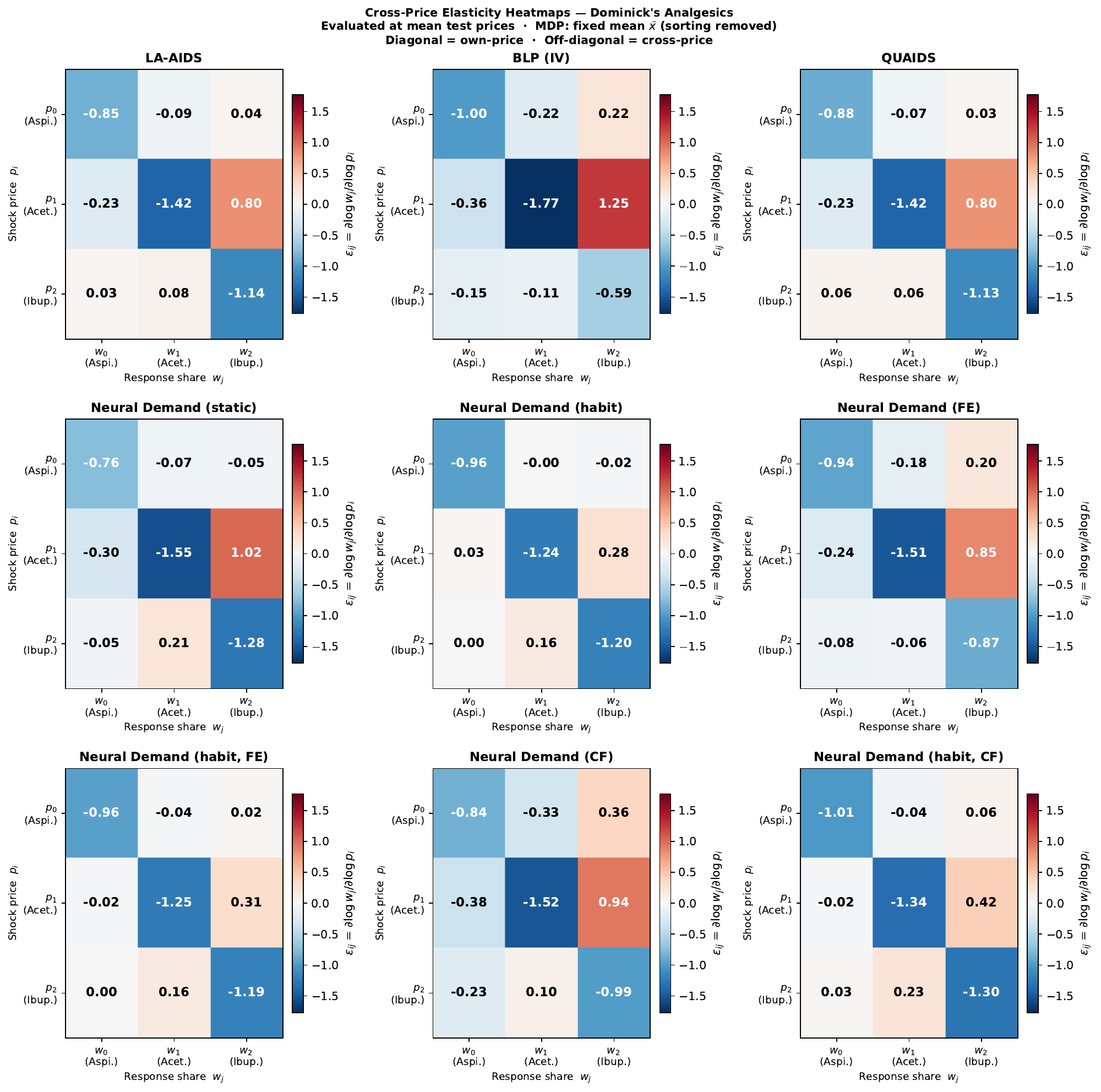}
  \caption{Cross-price elasticity matrices for seven key specifications --- Dominick's Analgesics. Evaluated at mean test prices; habit stock fixed at cross-sectional mean (MDP, sorting removed). Diagonal: own-price quantity elasticities. Off-diagonal: cross-price quantity elasticities. The aspirin--ibuprofen cross-price effects (row 1 column 3; row 3 column 1) converge toward zero under all neural demand specifications and are essentially zero under the habit and habit, FE models. The acetaminophen--ibuprofen cross (row 2 column 3; row 3 column 2) is robustly positive across all specifications. The full eleven-specification matrix is reported in Appendix Figure~\ref{fig:dom_cross_elast_full}.}
  \label{fig:dom_cross_elast}
\end{figure}

Table~\ref{tab:dom_elast} reports own-price \emph{quantity} elasticities evaluated at mean test prices,
and Figure~\ref{fig:dom_cross_elast} reports the corresponding $3\times 3$ cross-price matrices (with
the habit stock fixed at its cross-sectional mean to remove state selection). 

\begin{table}[htbp]
  \centering
  \caption{Own-Price Quantity Elasticities --- Dominick's Analgesics (5 runs; mean $\pm$ std)}
  \label{tab:dom_elast}
  \begin{threeparttable}
     \begin{tabular}{lccc}
      \toprule
        \textbf{Model} & \textbf{$\epsilon_{00}$ (ASP)} & \textbf{$\epsilon_{11}$ (ACET)} & \textbf{$\epsilon_{22}$ (IBU)} \\
        \midrule
        LA-AIDS & $-0.849 $ & $-1.420 $ & $-1.140 $ \\
        BLP (IV) & $-0.998 $ & $-1.769 $ & $-0.587 $ \\
        QUAIDS & $-0.882 $ & $-1.423 $ & $-1.129 $ \\
        Series Est. & $-0.708 $ & $-1.629 $ & $-1.224 $ \\
        Linear Demand (Shared) & $-0.163 $ & $-0.215 $ & $-0.190 $ \\
        Linear Demand (GoodSpec) & $-0.164 $ & $-0.205 $ & $-0.181 $ \\
        Linear Demand (Orth) & $-1.165 $ & $-1.125 $ & $-1.177 $ \\
        Neural Demand (window) & $-0.980 \pm 0.007$ & $-1.186 \pm 0.003$ & $-1.436 \pm 0.009$ \\
        Neural Demand (static) & $-0.757 \pm 0.033$ & $-1.552 \pm 0.035$ & $-1.284 \pm 0.012$ \\
        Neural Demand (habit) & $-0.958 \pm 0.005$ & $-1.237 \pm 0.011$ & $-1.201 \pm 0.018$ \\
        Neural Demand (FE) & $-0.937 \pm 0.023$ & $-1.506 \pm 0.037$ & $-0.872 \pm 0.037$ \\
        Neural Demand (habit, FE) & $-0.956 \pm 0.006$ & $-1.248 \pm 0.013$ & $-1.189 \pm 0.012$ \\
        Neural Demand (CF) & $-0.845 \pm 0.024$ & $-1.518 \pm 0.026$ & $-0.995 \pm 0.025$ \\
        Neural Demand (habit, CF) & $-1.012 \pm 0.012$ & $-1.341 \pm 0.013$ & $-1.296 \pm 0.010$ \\
        Neural Demand (habit, FE, CF) & $-0.962 \pm 0.011$ & $-1.318 \pm 0.018$ & $-1.270 \pm 0.014$ \\
        PlaceboND & $-0.779 \pm 0.037$ & $-1.468 \pm 0.027$ & $-1.318 \pm 0.018$ \\
        \bottomrule
    \end{tabular}
    \begin{tablenotes}\small
      \item Numerical own-price quantity elasticities at mean test prices. For habit models, habit stock fixed at cross-sectional mean (sorting removed).
    \end{tablenotes}
  \end{threeparttable}
\end{table}

We draw three main conclusions. First, the habit-augmented neural specifications deliver a stable and economically plausible own-price
profile. Across the habit variants, aspirin is close to unit elastic ($\epsilon_{00}\approx-0.96$ to $-1.01$),
acetaminophen is more elastic ($\epsilon_{11}\approx-1.24$ to $-1.34$), and ibuprofen is moderately
elastic ($\epsilon_{22}\approx-1.19$ to $-1.30$). The small run-to-run dispersion indicates that these
elasticities are not an artifact of initialization. By contrast, purely linear share systems (Linear Demand
Shared/GoodSpec) imply implausibly small elasticities near zero, while BLP (IV) yields a markedly less
elastic ibuprofen response ($-0.59$), highlighting the sensitivity of implied own-price effects to
functional-form restrictions in this category.

Second, moving from the static neural model to the habit model increases the magnitude of the aspirin
own-price elasticity (from $-0.76$ to about $-0.96$). This is consistent with the decomposition logic in
Section~\ref{sec:demand_decomp}: omitting the state variable compresses estimated price responses when
price is correlated with persistent demand conditions. Conditioning on the habit/state variable and
evaluating elasticities at a fixed state (sorting removed) yields a more negative, and more stable,
aspirin elasticity. The placebo model, which breaks the temporal link while preserving the marginal
distribution of history inputs, produces elasticities close to the static model rather than the habit
model, reinforcing that the shift is driven by meaningful state dependence rather than generic
regularization or extra covariates.

Third, the cross-price matrices in Figure~\ref{fig:dom_cross_elast} indicate that the principal structural
substitution channel in this category is between acetaminophen and ibuprofen. Across specifications,
the acetaminophen--ibuprofen cross effects are positive and sizable, whereas the aspirin--ibuprofen
cross effects collapse toward zero under the habit specifications with fixed state. This pattern aligns
with the demand decomposition: once persistent state/segment correlation is controlled for, there is
little evidence of meaningful substitution between aspirin and ibuprofen at the category level, while
substitution between the two broad-spectrum analgesics remains robust.

Finally, the control-function correction mainly affects elasticities through causal interpretation rather
than predictive fit. Relative to the static neural model, the CF variant modestly changes the implied
own-price elasticities; the habit+CF variants remain close to the habit baseline, suggesting that much
of the persistent price--share correlation is already absorbed by the state variable in this application.

\subsubsection{Demand Decomposition: Structural Price Effects versus State--Segment Correlation}
\label{sec:demand_decomp}

\begin{figure}[htbp]
  \centering
  \includegraphics[width=0.6\linewidth]{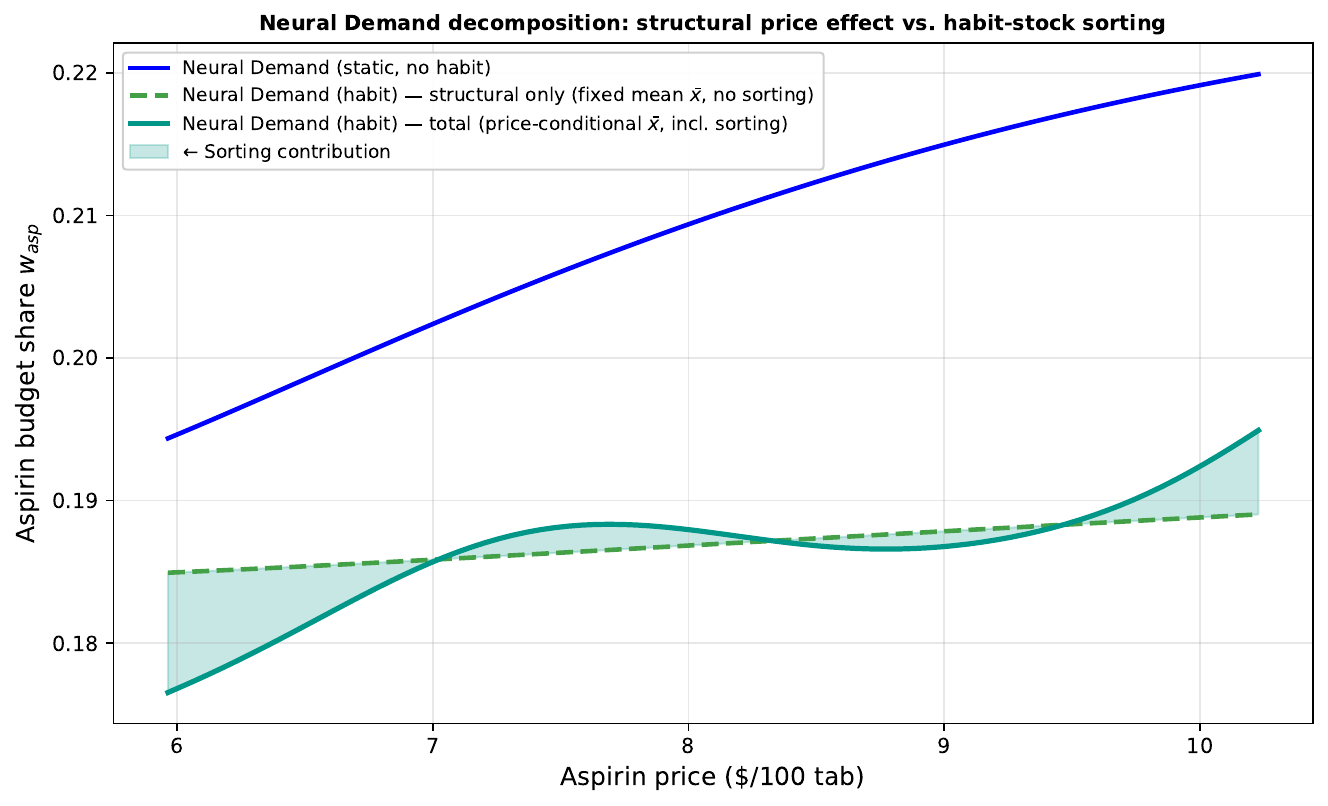}
  \caption{Neural demand decomposition: structural price effect vs.\ state--segment correlation --- Dominick's Analgesics. \textit{Blue}: Neural Demand (static), plotting the model-implied conditional mean aspirin budget share as aspirin price varies. \textit{Dashed green}: Neural Demand (habit) evaluated at a fixed state (habit stock set to the cross-sectional mean $\bar x$), isolating the structural price effect holding the state constant. \textit{Solid teal}: Neural Demand (habit) evaluated at the price-conditional state distribution (allowing $\bar x$ to vary with price as in the data), combining the structural effect with state selection. \textit{Shaded}: contribution of state--segment correlation (gap between teal and dashed green).}
  \label{fig:dom_decomposition}
\end{figure}

Figure~\ref{fig:dom_decomposition} summarizes the key mechanism in the Dominick's application:
the strong upward-sloping aspirin \emph{price--share} relationship in static models is a reduced-form
correlation driven by persistent heterogeneity, not a structural price effect.\footnote{
The decomposition is in \emph{shares}. A nearly flat share response is compatible with a substantially
downward-sloping quantity response because $w_i=p_i q_i/y$ implies
$\partial\ln w_i/\partial\ln p_i = 1 + \partial\ln q_i/\partial\ln p_i$ holding $y$ fixed. Consistent with
this, Table~\ref{tab:dom_elast} reports aspirin own-price \emph{quantity} elasticities close to $-1$ under
the habit specifications, so the implied share elasticity is near zero.} The static neural model
(blue) predicts higher aspirin shares at higher aspirin prices. The habit/state-augmented model allows
this relationship to be decomposed by holding the state fixed. Evaluating the habit model at a fixed
habit stock (dashed green, $\bar x=\mathbb E[\bar x]$) yields an almost flat \emph{share} response, while
evaluating at the price-conditional habit stock (solid teal, $\bar x=\mathbb E[\bar x\mid p_{\mathrm{asp}}]$)
partially restores the upward slope. The shaded region is therefore the contribution of
state--price correlation (selection into different store segments/states) to the overall price--share
relationship.

\begin{figure}[htbp]
  \centering
  \includegraphics[width=\linewidth]{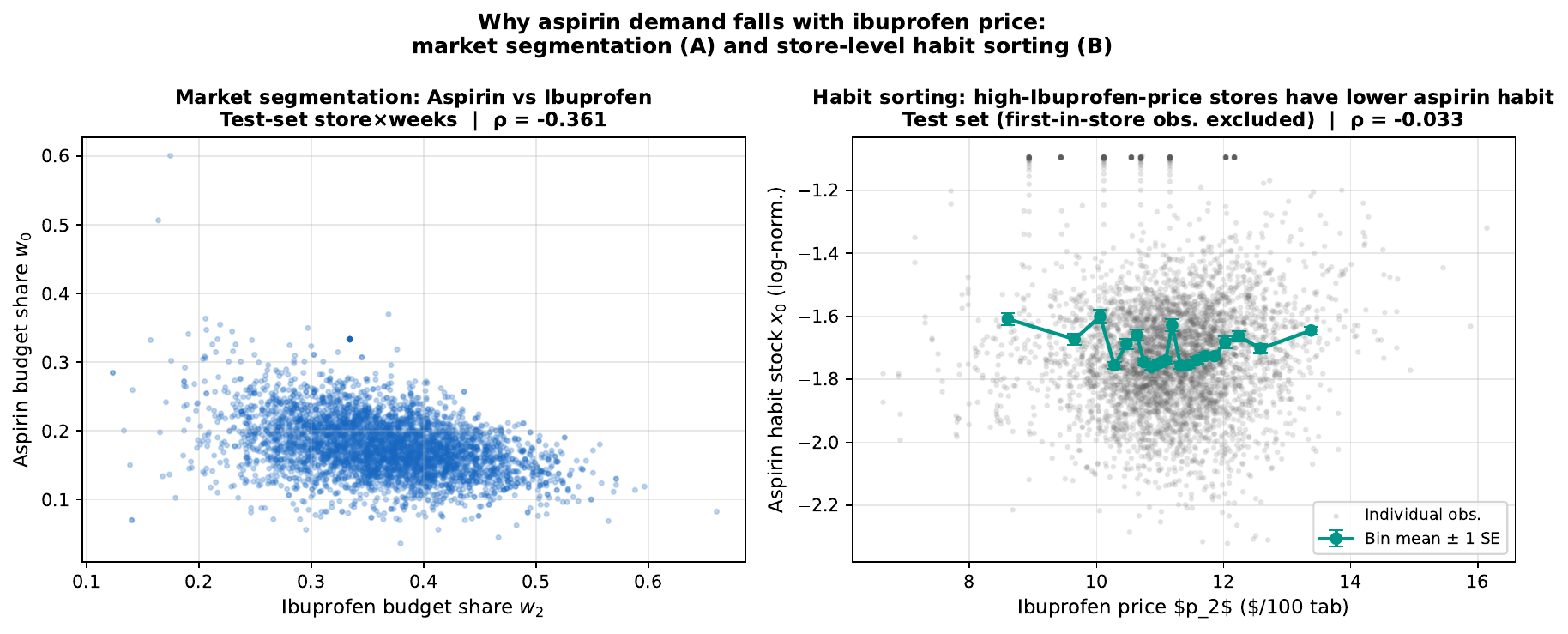}
  \caption{Market segmentation and state selection --- Dominick's Analgesics, test set. \textit{Left:}
  Aspirin budget share $w_0$ vs.\ ibuprofen budget share $w_2$ (Spearman $\rho=-0.361$), consistent
  with persistent segmentation across stores. \textit{Right:} Aspirin habit stock $\bar x_0$ vs.\ ibuprofen
  price (Spearman $\rho=-0.033$; binned mean $\pm 1$ SE), indicating little cross-good state selection
  with respect to ibuprofen price.}
  \label{fig:dom_segmentation}
\end{figure}

Figure~\ref{fig:dom_segmentation} indicates that persistent segmentation is an important feature of
the category. Panel~A shows a strong negative association between aspirin and ibuprofen budget
shares across test-set store-weeks (Spearman $\rho=-0.361$), consistent with stable store-level
differences in the composition of analgesic demand. Panel~B shows essentially no relationship between
the aspirin state (habit stock) and the ibuprofen price (Spearman $\rho=-0.033$), suggesting that the
confounding operates primarily through persistent segmentation rather than cross-good ``habit sorting''
with respect to ibuprofen prices.

\begin{figure}[htbp]
  \centering
  \includegraphics[width=0.8\linewidth]{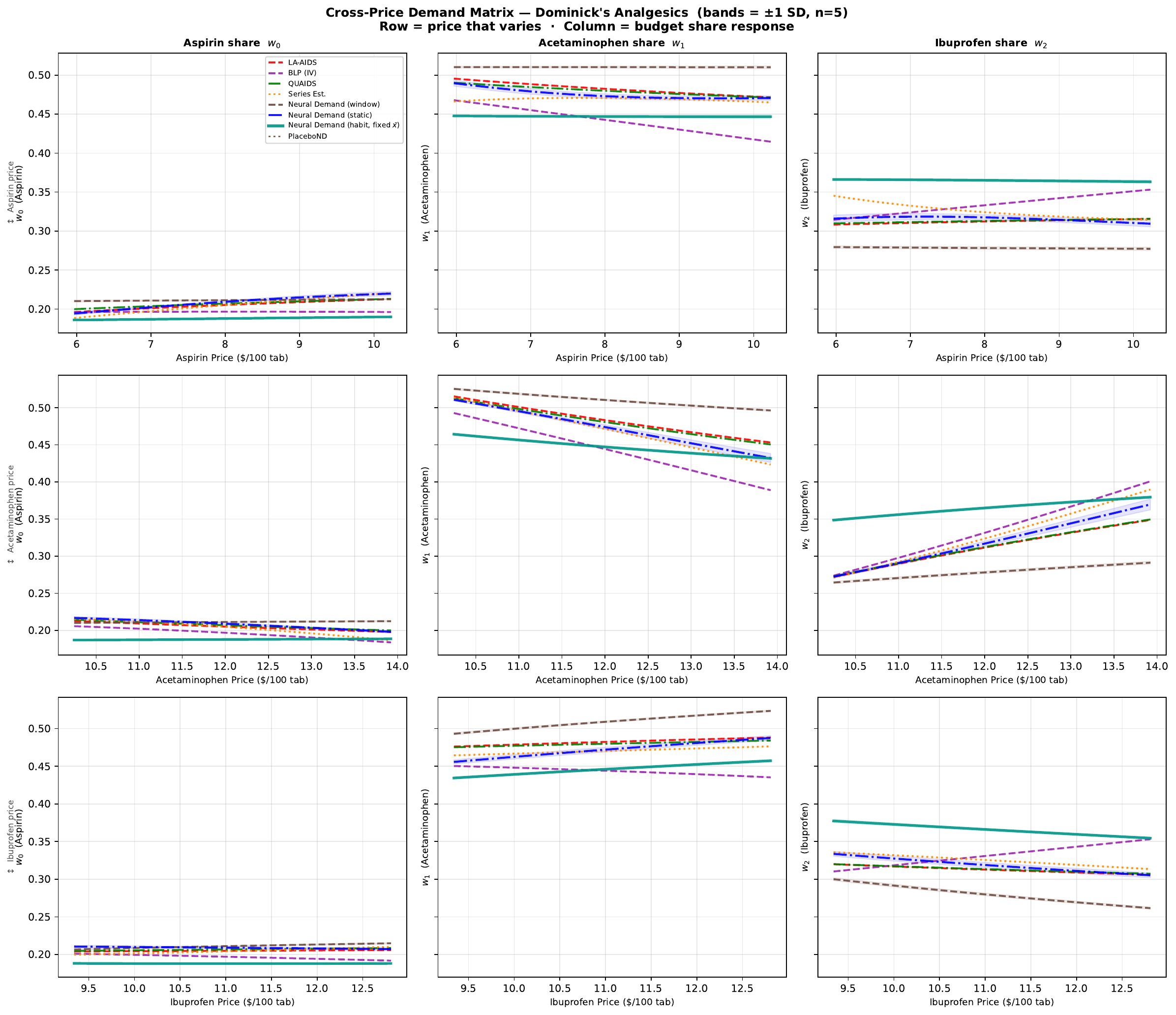}
  \caption{Cross-price demand matrix --- Dominick's Analgesics ($\pm 1$ SD, $n=5$). Row = price that
  varies; column = budget share response. All other prices and income are held at evaluation-sample
  means. Neural Demand (habit, fixed $\bar x$) evaluates the habit model at $\bar x=\mathbb E[\bar x]$ to
  remove state selection. The aspirin--ibuprofen cross-price slopes largely vanish under the fixed-state
  habit specification, while acetaminophen--ibuprofen substitution remains robustly positive.}
  \label{fig:dom_demand_matrix}
\end{figure}

Figure~\ref{fig:dom_demand_matrix} reports cross-price responses (prices varied over $\pm1$ SD with
other prices and expenditure held at test means). The salient comparison is the aspirin--ibuprofen
relationship: static and parametric benchmarks imply positive cross-effects, whereas the habit model
evaluated at fixed $\bar x$ yields cross-responses close to zero, indicating that apparent
aspirin--ibuprofen substitution largely reflects state/segment correlation. In contrast, the
acetaminophen--ibuprofen cross-effects remain positive across specifications, consistent with genuine
substitution between the two broad-spectrum analgesics.

Therefore, the pronounced aspirin price--share gradients produced by static specifications are
primarily a manifestation of state--segment correlation rather than structural substitution. Holding the
habit/state fixed yields a near-flat \emph{share} response for aspirin (consistent with a quantity
elasticity close to $-1$) and collapses the apparent aspirin--ibuprofen cross effect, while
acetaminophen--ibuprofen substitution remains robust. The implication is that, in repeat-purchase
categories with persistent heterogeneity, static demand systems can conflate selection with
substitution and thereby distort welfare calculations from price changes; incorporating state dependence
provides a disciplined way to separate these channels.

\subsubsection{Welfare Analysis: 10\% Ibuprofen Price Increase}
\label{sec:dom_welfare}

Table~\ref{tab:dom_welfare_ibuprofen} reports compensating-variation (CV) losses from a 10\% increase
in the ibuprofen price, computed by a 100-step Riemann approximation to the line integral. Because
expenditure is proxied by within-category revenue, the CV magnitudes are interpreted as
category-expenditure equivalents and comparisons across models are most naturally read as ordinal
(Section~\ref{sec:identification}). Appendix~\ref{app:regularity_dashboard} documents that the welfare
computations are based on demand systems that are close to integrable on the evaluation support, with
habit-augmented models exhibiting substantially smaller symmetry and homogeneity deviations. We report welfare effects due to similar changes in the price of Aspirin and Acetaminophen in Appendix \ref{sec:additional-dominicks-welfare}.

\begin{table}[htbp]
  \centering
  \caption{Welfare Loss from 10\% Ibuprofen Price Increase --- Dominick\'s Analgesics (5 run(s); mean $\pm$ std)}
  \label{tab:dom_welfare_ibuprofen}
  \begin{threeparttable}
    \begin{tabular}{lcr}
      \toprule
      \textbf{Model} & \textbf{CV Loss (\$)} & \textbf{vs Neural Demand (static)} \\
      \midrule
      LA-AIDS & $-34.8710 $ & +1.1\% \\
      BLP (IV) & $-38.1498 $ & -8.2\% \\
      QUAIDS & $-34.9710 $ & +0.8\% \\
      Series Est. & $-36.1016 $ & -2.4\% \\
      Linear Demand (Shared) & $-42.1538 $ & -19.6\% \\
      Linear Demand (GoodSpec) & $-42.1721 $ & -19.6\% \\
      Linear Demand (Orth) & $-31.0624 $ & +11.9\% \\
      Neural Demand (static) & $-35.2502 \pm 0.3796$ &  \\
      Neural Demand (window) & $-30.6534 \pm 0.1800$ & +13.0\% \\
      Neural Demand (habit) & $-40.6668 \pm 0.1772$ & -15.4\% \\
      Neural Demand (FE) & $-37.4868 \pm 1.0057$ & -6.3\% \\
      Neural Demand (habit, FE) & $-40.9997 \pm 0.1543$ & -16.3\% \\
      Neural Demand (CF) & $-37.7469 \pm 0.2669$ & -7.1\% \\
      Neural Demand (habit, CF) & $-40.8591 \pm 0.1217$ & -15.9\% \\
      Neural Demand (habit, FE, CF) & $-40.8201 \pm 0.1489$ & -15.8\% \\
      PlaceboND & $-34.6134 \pm 0.3288$ & +1.8\% \\
      \bottomrule
    \end{tabular}
    \begin{tablenotes}\small
      \item Compensating variation via 100-step Riemann sum, $p_{Ibuprofen}\to(1+0.1)\,p_{Ibuprofen}$.
    \end{tablenotes}
  \end{threeparttable}
\end{table}

Three patterns emerge. First, welfare conclusions are sensitive to whether demand is modeled as
state dependent. Relative to the static neural demand system (CV loss $-35.25$), the habit-augmented
variants imply markedly larger welfare losses, around $-40.7$ to $-41.0$, a difference of
approximately 15--16\%. This gap persists in the fixed-effect and control-function variants, indicating
that the welfare impact of state dependence is not driven solely by time-invariant store heterogeneity
or by the endogenous component of price variation. The placebo model, which breaks the temporal
link in the history inputs while preserving their marginal distribution, yields CV close to the static
model, reinforcing that the welfare difference is tied to correctly aligned dynamics rather than to a
generic ``more inputs'' effect.

Second, the habit-induced increase in welfare loss is economically consistent with the elasticity and
substitution results. The habit specifications imply stronger persistence in brand/category allocation
and attenuated substitution patterns once the state is held fixed, so the same ibuprofen price increase
generates a larger compensating payment to restore baseline utility. In contrast, models that impose
more restrictive substitution patterns deliver substantially different welfare magnitudes: the linear
shared/goodspecific systems imply the largest losses (about $-42.17$), while some flexible-history
specifications (e.g., the window model) imply noticeably smaller losses.

Third, the static parametric share systems (LA-AIDS and QUAIDS) deliver welfare losses close to the
static neural benchmark (within about 1\%), while BLP and the linear demand systems deliver larger
deviations. Overall, the main empirical implication of Table~\ref{tab:dom_welfare_ibuprofen} is that
incorporating state dependence materially affects welfare conclusions in this repeat-purchase category:
relative to static specifications, habit/state augmentation implies meaningfully larger welfare
losses from price increases, even when evaluated in the same expenditure units and under the same
price shock.

\section{Conclusion}
\label{sec:conclusion}

This paper argues that state dependence is a first-order feature of demand in repeat-purchase markets
and that treating demand as static can change both substitution patterns and welfare conclusions from
price variation. The core issue is that static demand systems, even when flexible, tend to load
persistent heterogeneity and purchase-history effects onto contemporaneous prices and expenditure,
thereby conflating selection with substitution.

We address this with a demand system for continuous budget allocation designed for scanner-style
share data. The estimator targets the conditional mean share mapping on the simplex by minimizing
Kullback--Leibler divergence between predicted and observed budget shares. Shares are parameterized
through a softmax map (enforcing adding-up by construction) and disciplined with economic
regularization and transparent near-integrability diagnostics to support welfare analysis. The resulting
mapping is differentiable, enabling computation of elasticities, Slutsky objects, and compensating
variation. State dependence enters through a low-dimensional habit stock constructed as an EWMA of
past consumption. When relevant, we also implement a control-function correction for endogenous
prices and evaluate counterfactual demand at the purged residual.

A sequence of simulation designs validates the approach. Under CES preferences, the estimator
recovers demand curves, the full cross-price elasticity matrix, and compensating variation essentially
exactly. More importantly, the simulations separate functional-form limitations from economic
structure. In static DGPs, flexible estimators perform well by approximating curvature and kinks. In
the habit-formation DGP, however, all static estimators---including flexible nonparametric ones---fail
in the same way, because the dominant source of variation is omitted state. Adding the habit stock
produces a discrete improvement (large reductions in RMSE and KL). In the endogenous-price DGP,
the control-function correction is essential for counterfactual accuracy. Profiling the habit-decay
parameter yields a set-valued conclusion: short memory is rejected, while a range of long-memory
values is observationally near-equivalent, motivating reporting welfare either at $\hat\delta$ and/or over
the identified set.

In the Dominick's analgesics application, state augmentation reduces out-of-sample prediction error by
about one third relative to standard share systems, and a shuffled-history placebo confirms that the
gain is not a generic ``more inputs'' effect. Substantively, the implied substitution patterns change:
static specifications generate steep aspirin price--share gradients and an apparent aspirin--ibuprofen
cross-price effect, whereas conditioning on the habit/state and holding it fixed collapses the
aspirin--ibuprofen cross effect toward zero while preserving robust acetaminophen--ibuprofen
substitution. These differences translate into welfare. For a 10\% ibuprofen price increase, the habit
specification implies compensating-variation losses about 15--16\% larger than the static neural model.
Near-integrability audits show that habit specifications also exhibit substantially smaller symmetry and
homogeneity deviations, with curvature violations that are frequent but quantitatively small on the
evaluation support.

Several extensions are natural. Stronger enforcement of economic regularity (including explicit
homogeneity and curvature constraints) would sharpen welfare interpretation when exact
rationalizability is required. Developing formal inference for elasticities and welfare functionals in
regularized state-dependent demand systems remains an important direction, particularly under weak
identification of persistence parameters. Finally, richer state representations (inventory, promotions,
multi-category linkages) and household-level panels would broaden the policy reach of the approach
for applications such as inflation measurement, tax incidence, and counterfactual evaluation of pricing
interventions.


\bibliographystyle{apalike}
\bibliography{bib.bib}

\appendix

\section{Further Simulation Results}

\subsection{Training Convergence by DGP}

\begin{figure}[htbp]
    \centering
    \includegraphics[width=\linewidth]{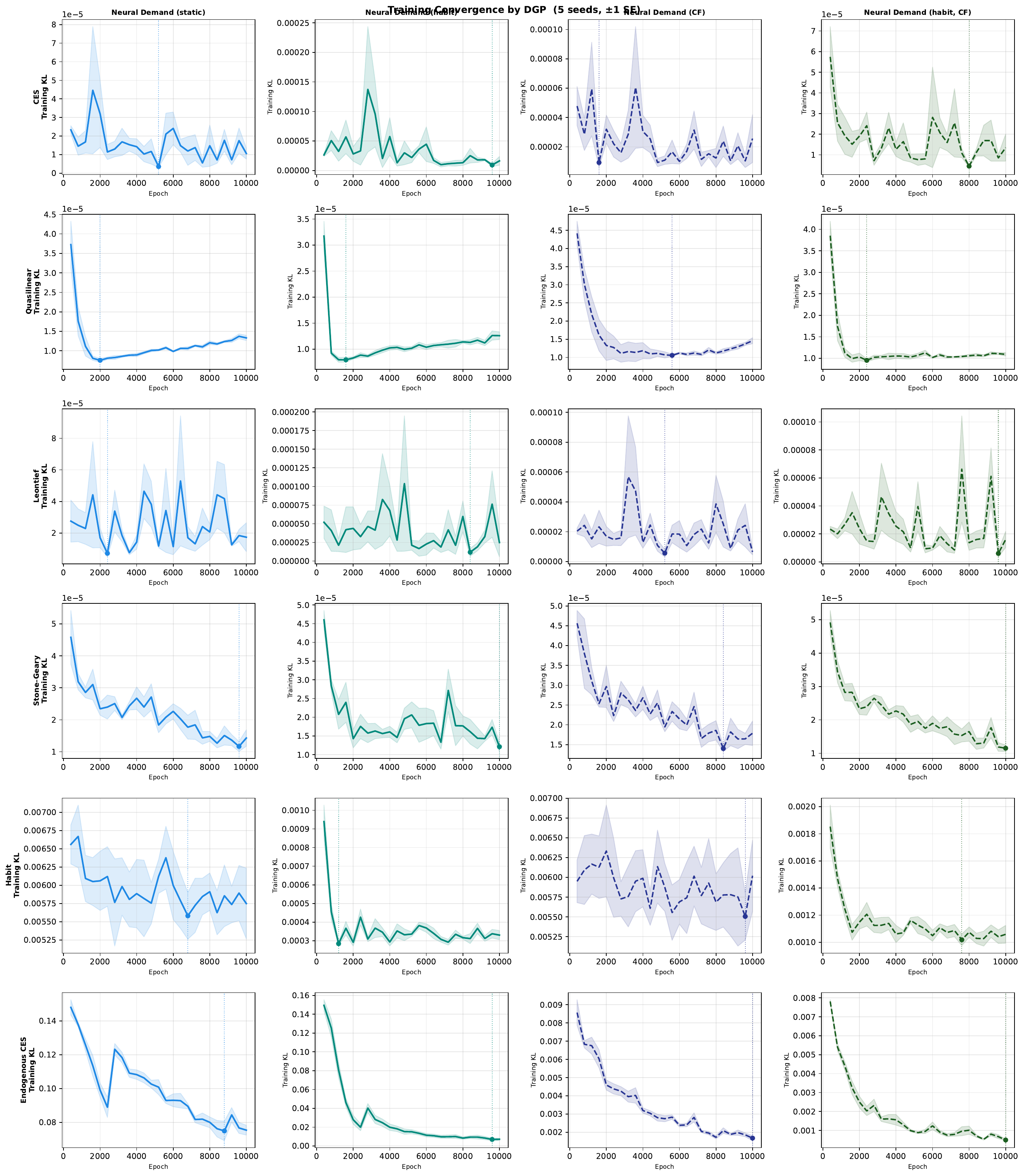}
    \caption{Training KL divergence by DGP across epochs ($\pm$1 SE, 5 seeds).
    Each row corresponds to a specific Data Generating Process (DGP), and each column represents a distinct ND architecture: \textit{Static}, \textit{Habit-persistent}, \textit{Control Function (CF)}, and the integrated \textit{Habit-CF} model. 
    The y-axis displays the Training Kullback-Leibler (KL) divergence over 10,000 epochs, illustrating the optimization trajectory for each estimator. 
    Solid and dashed lines represent the mean training loss across 5 independent seeds, while the shaded regions denote $\pm 1$ standard error. 
    Vertical dotted lines and markers indicate the checkpoint corresponding to the minimum training loss. }
    \label{fig:sim_convergence}
\end{figure}

Figure~\ref{fig:sim_convergence} illustrates the training convergence of the Neural Demand (ND) model variants across all simulated Data Generating Processes (DGPs). The plots show the mean training KL divergence over 5 independent runs, with shaded areas representing $\pm 1$ standard error. 

Overall, the ND architectures demonstrate robust convergence across diverse functional forms. We observe that the \textit{Endogenous CES} DGP is the most computationally demanding, resulting in the highest terminal KL divergence, likely due to the increased complexity of the price-endogeneity mapping. Notably, for the \textit{Habit} DGP, models incorporating habit-persistent layers (ND habit and ND habit, CF) achieve a terminal loss approximately one order of magnitude lower than static specifications. This confirms that the neural network effectively leverages the temporal structure when appropriately specified. While DGPs such as \textit{Leontief} and \textit{CES} exhibit higher variance and more frequent oscillations during the optimization process, all models reach a stable plateau by 10,000 epochs, justifying the choice of training duration.

\section{Further Empirical Results}

\subsection{First-Stage Diagnostics}
\label{sec:first_stage}

Table~\ref{tab:first_stage_fstats} reports relevance diagnostics for the Hausman-style instruments used
in the control function procedure (Section~\ref{sec:neural_iv}). For each good $g\in\{1,\ldots,G\}$, we
estimate the first-stage projection
\[
\ln p_{git} = Z_{it}^\top \pi_g + v_{git},
\]
where $Z_{it}$ stacks the excluded instruments (revenue-weighted mean \emph{log} prices of each good
across all stores other than $i$ in week $t$). We report: (i) the first-stage $F$-statistic for joint
significance of the excluded instruments, (ii) the overall $R^2$, and (iii) the partial $R^2$ attributable
to the excluded instruments.

\begin{table}[htbp]
  \centering
  \caption{First-Stage Diagnostics --- Hausman Instruments (Dominick's Analgesics)}
  \label{tab:first_stage_fstats}
  \begin{tabular}{lccc}
    \toprule
    \textbf{Price equation} & \textbf{First-stage $F$} & \textbf{First-stage $R^2$} & \textbf{Partial $R^2$} \\
    \midrule
    Aspirin & 43285.80 & 0.645 & 0.603 \\
    Acetaminophen & 49300.16 & 0.664 & 0.633 \\
    Ibuprofen & 35854.14 & 0.619 & 0.557 \\
    \bottomrule
  \end{tabular}
  \caption*{\footnotesize OLS first-stage regression of $\ln p_{git}$ on the full set of Hausman
  instruments. The $F$-statistic tests joint significance of the excluded instruments; values below 10
  indicate weak instruments \citep{staiger1994instrumental}. The partial $R^2$ is the incremental
  explanatory power of the excluded instruments relative to a first stage with included controls only.}
\end{table}

The first-stage $F$-statistics are extremely large for all three price equations, far exceeding
conventional weak-IV thresholds \citep{staiger1994instrumental}. The partial $R^2$ values indicate
that the excluded instruments explain a substantial fraction of within-sample log-price variation,
consistent with common price movements across stores within a week (e.g., network-wide cost shocks).
These diagnostics address the \emph{relevance} condition. The \emph{exclusion} restriction remains the
identifying assumption and is discussed in Section~\ref{sec:neural_iv}. Because the control-function
approach uses projection residuals rather than a single 2SLS coefficient, we emphasize
equation-by-equation first-stage strength (the $F$-statistic and partial $R^2$) as the primary weak-IV
check.

\subsection{Cross-Price Elasticity Heatmap} \label{sec:cp-elas-full}

\begin{figure}[htbp]
    \centering
    \includegraphics[width=\linewidth]{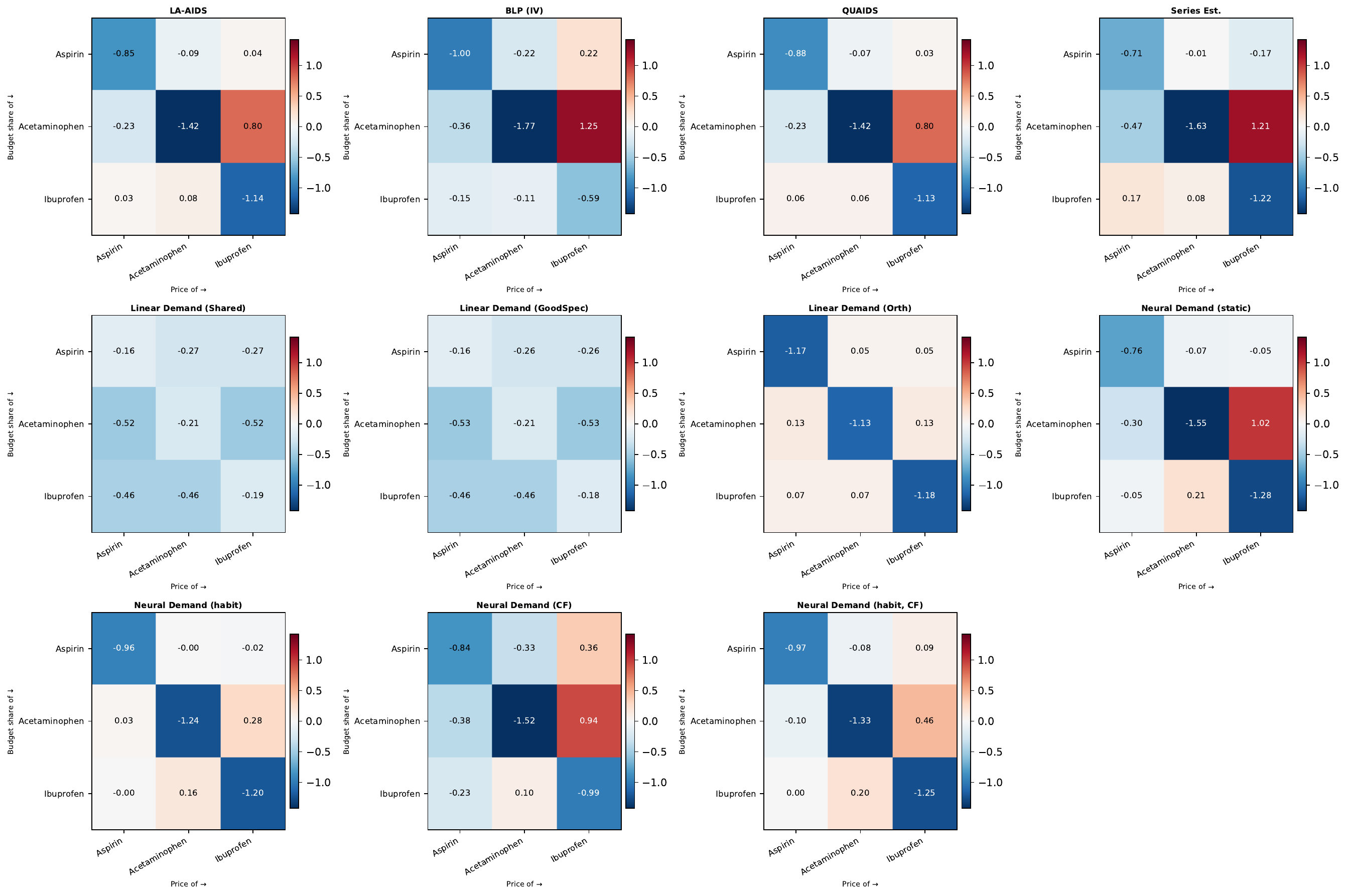}
    \caption{Full $3 \times 3$ cross-price elasticity heatmaps for the control function specifications --- Dominick's Analgesics. Rows: Neural Demand (CF) and Neural Demand (habit, CF), evaluated at $\hat{v}_t = 0$ (causal demand function). }
    \label{fig:dom_cross_elast_full}
\end{figure}

Figure~\ref{fig:dom_cross_elast_full} reports the full $3\times 3$ price elasticity matrices across
specifications, evaluated at mean test covariates and, for habit models, with the habit stock fixed at
its cross-sectional mean to remove state selection. 

Across models, own-price elasticities (diagonal)
are negative and of economically plausible magnitude, while cross-price responses (off-diagonals)
reveal a systematic pattern: substitution between acetaminophen and ibuprofen is robustly positive in
nearly all specifications, whereas the aspirin--ibuprofen cross effects are small and, in the habit
specifications, essentially zero. In particular, the habit-augmented neural models largely eliminate
the apparent aspirin--ibuprofen substitution present in several static/parametric benchmarks, while
preserving strong acetaminophen--ibuprofen substitution. This supports the interpretation that a
substantial share of the static aspirin--ibuprofen cross-price relationship reflects persistent
state/segment correlation rather than structural substitution.

\subsection{Training Convergence}

\begin{figure}[htbp]
    \centering
    \includegraphics[width=0.75\linewidth]{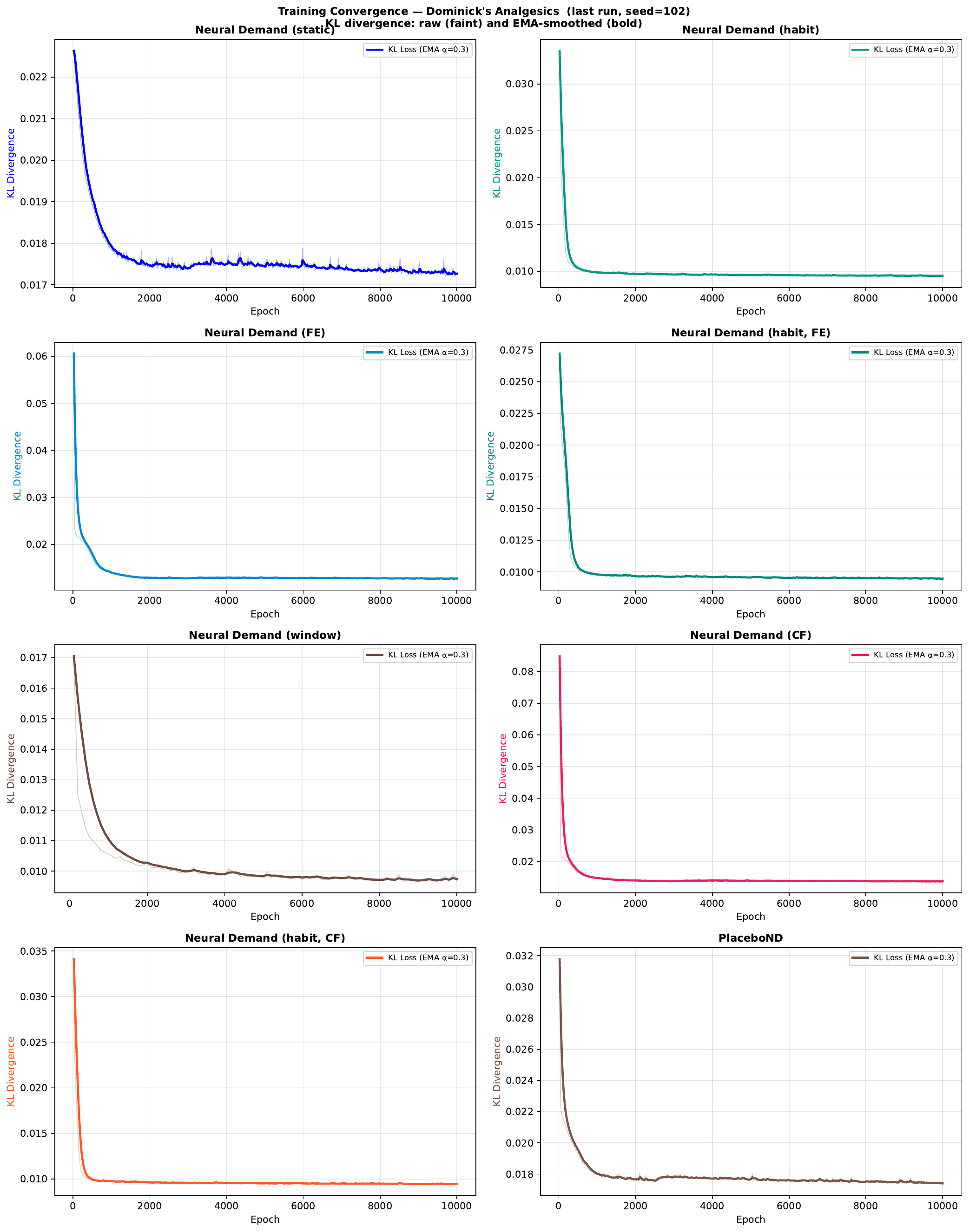}
    \caption{Training KL divergence on the Dominick's analgesics panel ($\pm$1 SE, 5 seeds) for the neural demand models over 10,000 epochs. }
    \label{fig:dom_training_convergence}
\end{figure}

Figure~\ref{fig:dom_training_convergence} shows training KL divergence across epochs for the static and habit-augmented models. The static model (trained for 3,000 epochs) converges smoothly within approximately 1,000--1,500 epochs. The habit-augmented model (trained for 10,000 epochs) requires approximately 2,000--3,000 epochs to fully exploit the habit state, reflecting the additional complexity of learning how the EWMA habit stock interacts with prices and income. Both models are trained to the best checkpoint by full-dataset KL evaluated every 50 epochs, as described in Section~\ref{sec:nds}.

\subsection{Near-Integrability Diagnostics}
\label{app:regularity_dashboard}

This appendix reports diagnostic measures that summarize the extent to which the estimated share
systems satisfy (or approximately satisfy) the integrability conditions underlying welfare analysis on the Dominicks data.
The goal is \emph{not} to claim exact rationalizability of a flexible estimator, but to transparently
quantify \emph{near-integrability} achieved by economic regularization and to provide readers with
a compact audit of the objects entering our welfare computations.

\subsubsection{Diagnostic definitions}

Let $\hat w(p,y,\bar x)$ denote the fitted budget share mapping with $\hat w(\cdot)\in\Delta^{G-1}$.
Let $\hat q_i(p,y,\bar x)=\hat w_i(p,y,\bar x)\,y/p_i$ denote the implied Marshallian quantities at
fixed $(y,\bar x)$. Let $\hat S(p,y,\bar x)$ denote the implied Slutsky matrix. For each observation in an evaluation
sample, we compute:

\paragraph{Adding-up.}
Although adding-up holds by construction of the softmax output map, we report the numerical error
\[
D_{\mathrm{add}} \equiv \bigl|\mathbf 1^\top \hat w - 1\bigr|.
\]

\paragraph{Slutsky symmetry.}
We summarize symmetry deviations using the Frobenius norm:
\[
D_{\mathrm{sym}} \equiv \|\hat S-\hat S^\top\|_F.
\]
Table~\ref{tab:regularity_dashboard} reports the mean and maximum of $D_{\mathrm{sym}}$ over the
evaluation sample.

\paragraph{Curvature / negativity diagnostics.}
Let $\tilde S\equiv(\hat S+\hat S^\top)/2$ be the symmetrized Slutsky matrix. We report two
diagnostics: (i) the incidence of positive curvature,
\[
\mathbbm{1}\{\lambda_{\max}(\tilde S)>0\},
\]
and (ii) the magnitude of positive curvature,
\[
D_{\mathrm{curv}} \equiv \max\{\lambda_{\max}(\tilde S),0\}.
\]
These measures are zero when $\tilde S$ is negative semidefinite and increase with violations.

\paragraph{Homogeneity.}
Homogeneity of degree zero in $(p,y)$ is not imposed as a hard restriction. We therefore evaluate a
simple scaling diagnostic. For prespecified scalars $c\in\mathcal C$ (e.g., $\mathcal C=\{0.8,1.2\}$),
define
\[
D_{\mathrm{hom}}(c)\equiv \|\hat w(cp,cy,\bar x)-\hat w(p,y,\bar x)\|_{\infty},
\]
and report $\mathbb E_c[D_{\mathrm{hom}}(c)]$ averaged over both observations and $c\in\mathcal C$.

\subsubsection{Regularity dashboard by specification}

Table~\ref{tab:regularity_dashboard} reports out-of-sample diagnostics for the principal specifications studied in
the paper (static vs habit; with vs without economic regularization; and, where relevant, fixed-effect
variants). For models indexed by the decay parameter $\delta$, we report diagnostics at $\hat\delta$ and, when relevant, ranges over $\widehat{\mathcal D}$.

\begin{table}[t]
\centering
\caption{Near-integrability diagnostics (``regularity dashboard'').}
\label{tab:regularity_dashboard}
\begin{tabular}{lcccccc}
\toprule
Model & Test KL & $\mathbb E[D_{\mathrm{sym}}]$ & $\max D_{\mathrm{sym}}$ & $\Pr(\lambda_{\max}(\tilde S)>0)$ & $\mathbb E[D_{\mathrm{curv}}]$ & $\mathbb E[D_{\mathrm{hom}}]$ \\
\midrule
Static (no reg) & 0.0192 & 0.0499 & 0.5395 & 0.937 & 0.0027 & 0.0576 \\
Static (reg) & 0.0201 & 0.0353 & 0.4358 & 0.937 & 0.0019 & 0.0487 \\
Habit (no reg) & 0.0114 & 0.0157 & 0.2463 & 0.931 & 0.0004 & 0.0186 \\
Habit (reg) & 0.0113 & 0.0078 & 0.1279 & 0.930 & 0.0001 & 0.0109 \\
\addlinespace
FE static (reg) & 0.0164 & 0.0438 & 0.2553 & 0.937 & 0.0043 & 0.0722 \\
FE habit (reg) & 0.0114 & 0.0107 & 0.1009 & 0.937 & 0.0003 & 0.0175 \\
\bottomrule
\end{tabular}
\end{table}

Three patterns are notable. First, habit augmentation improves both fit and near-integrability. Relative to the static models, the
habit specifications reduce held-out KL loss substantially (from about $0.019$ to about $0.011$) and
sharply reduce symmetry and homogeneity deviations. For example, $\mathbb E[D_{\mathrm{sym}}]$
falls from $0.0499$ (static, no regularization) to $0.0157$ (habit, no regularization) and further to
$0.0078$ (habit, with regularization). Likewise, homogeneity deviations drop from roughly $0.058$ in
the static model to $0.011$ under the habit+regularization specification, indicating that the habit model
is substantially closer to degree-zero homogeneity in $(p,y)$ on the observed support despite the
restriction not being imposed as a hard constraint.

Second, the symmetry penalty operates as intended with limited cost in predictive fit. Moving from
``static (no reg)'' to ``static (reg)'' reduces average and worst-case symmetry violations
($\mathbb E[D_{\mathrm{sym}}]$ from $0.0499$ to $0.0353$; $\max D_{\mathrm{sym}}$ from $0.5395$ to
$0.4358$), and modestly improves curvature and homogeneity diagnostics. In the habit specification,
regularization further reduces both average and maximum symmetry deviations and improves the
homogeneity diagnostic, while leaving test KL essentially unchanged.

Third, curvature violations are frequent but quantitatively small. The incidence measure
$\Pr(\lambda_{\max}(\tilde S)>0)$ is high across all specifications (around $0.93$), reflecting that exact
negative semidefiniteness is not enforced. However, the magnitude of violations is small, especially in
habit models: $\mathbb E[D_{\mathrm{curv}}]$ falls from $2.7\times 10^{-3}$ in the static model to
$1\times 10^{-4}$ in the habit+regularization model. Thus, while the fitted systems are not exactly
integrable, they are empirically close to symmetry and homogeneity on the evaluation support, and
their curvature departures are small in magnitude.

Finally, adding store fixed effects improves predictive fit in the static model (test KL $0.0164$) but does
not systematically improve the regularity diagnostics relative to the corresponding non-FE models; in
particular, the FE static model exhibits larger homogeneity deviations. In contrast, the FE habit model
retains the strong fit and near-integrability properties of the baseline habit specification.

\subsubsection{Habit-Decay Identification}
\label{sec:dom_delta}

\begin{figure}[htbp]
  \centering
  \includegraphics[width=0.70\linewidth]{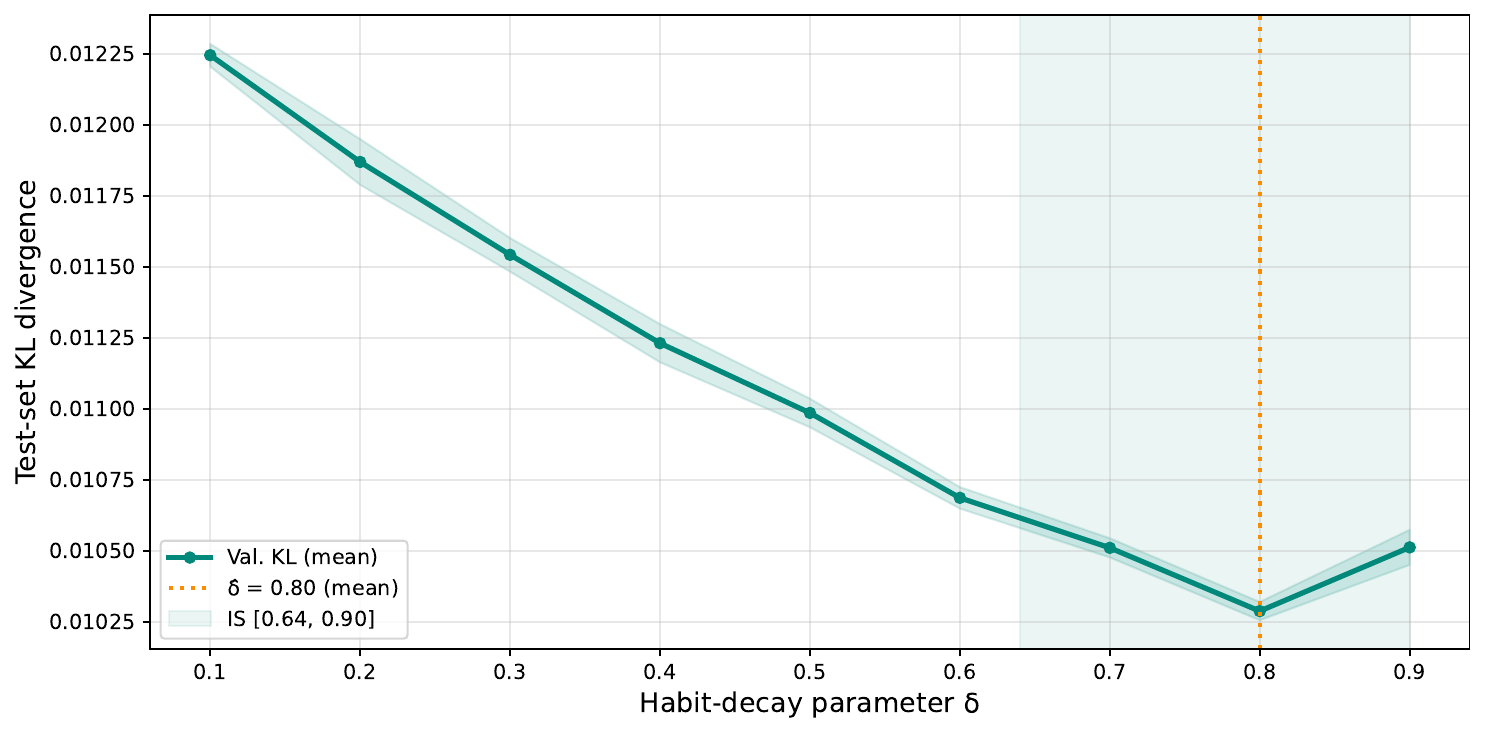}
  \caption{Profile KL criterion for $\delta$ --- Dominick's Analgesics ($\pm 1$ SD, 5 seeds). Minimum at $\hat\delta = 0.868$ (dotted vertical line). The identified set $[0.68, 0.90]$ (shaded) spans values within the 2-SD threshold. Values below $\delta = 0.5$ are sharply rejected; intermediate values are observationally near-equivalent.}
  \label{fig:dom_delta_profile}
\end{figure}

Table~\ref{tab:dom_delta} and Figure~\ref{fig:dom_delta_profile} summarize identification of the
habit-decay parameter $\delta$ using the out-of-sample (profile) KL criterion. 

\begin{table}[h]
  \centering
  \caption{Profile KL Identification of $\delta$ --- Dominick's Analgesics (5 seeds)}
  \label{tab:dom_delta}
   \begin{tabular}{lc}
        \toprule
    \textbf{Quantity} & \textbf{Value} \\
    \midrule
    $\hat{\delta}$ mean ± SE) & 0.800 ± 0.000 \\
    IS lo (mean) & 0.640 \\
    IS hi (mean) & 0.900 \\
    IS width (mean±SE) & 0.260 ± 0.024 \\
    \bottomrule
  \end{tabular}
\end{table}

The profile curve is
well behaved in the sense that very low decay values are rejected: test KL rises sharply for
$\delta\lesssim 0.5$, indicating that short-memory histories fit the data poorly. In contrast, the
criterion is relatively flat over a broad range of higher values, yielding a sizable identified set. Across
seeds, the profile minimum occurs at $\hat\delta\approx 0.80$, while the 2-SD identified set averages
$[0.64,0.90]$ with width about $0.26$. 

This pattern is consistent with weak (set) identification of
EWMA decay parameters in flexible state-dependent models: once the memory length is sufficiently
long, nearby values of $\delta$ generate highly collinear habit stocks on the observed support and are
therefore observationally near-equivalent in predictive performance. Practically, the profile results
pin down the conclusion that persistence is important and long-lived, while leaving the precise
half-life only partially identified; accordingly, we report substantive objects either at $\hat\delta$ or
over the identified set.

\subsection{Welfare} \label{sec:additional-dominicks-welfare}

\subsubsection{Aspirin Price increase}

Table~\ref{tab:dom_welfare_aspirin} shows that welfare rankings depend on whether demand is modeled
as state dependent. Relative to the static neural benchmark (CV $-23.76$), the habit-augmented
specifications imply \emph{smaller} losses (around $-20.9$ to $-21.8$, i.e.\ $+8$--$12\%$), consistent with
stronger persistence in the aspirin state and limited reallocation away from aspirin when its price rises.
By contrast, linear demand systems imply substantially larger losses (about $-28.2$), reflecting their
restrictive substitution patterns. The placebo model is close to the static benchmark, indicating that
the change in welfare is driven by correctly aligned history rather than by additional inputs.

\begin{table}[htbp]
  \centering
  \caption{Welfare Loss from 10\% Aspirin Price Increase --- Dominick\'s Analgesics (5 run(s); mean $\pm$ std)}
  \label{tab:dom_welfare_aspirin}
  \begin{threeparttable}
    \begin{tabular}{lcr}
       \toprule
      \textbf{Model} & \textbf{CV Loss (\$)} & \textbf{vs Neural Demand (static)} \\
      \midrule
      LA-AIDS & $-23.2182 $ & +2.3\% \\
      BLP (IV) & $-22.1176 $ & +6.9\% \\
      QUAIDS & $-23.3916 $ & +1.5\% \\
      Series Est. & $-23.3300 $ & +1.8\% \\
      Linear Demand (Shared) & $-28.2325 $ & -18.8\% \\
      Linear Demand (GoodSpec) & $-28.1783 $ & -18.6\% \\
      Linear Demand (Orth) & $-24.0706 $ & -1.3\% \\
      Neural Demand (static) & $-23.7586 \pm 0.1733$ &  \\
      Neural Demand (window) & $-23.7973 \pm 0.0663$ & -0.2\% \\
      Neural Demand (habit) & $-21.1675 \pm 0.0872$ & +10.9\% \\
      Neural Demand (FE) & $-24.2650 \pm 0.2513$ & -2.1\% \\
      Neural Demand (habit, FE) & $-21.6053 \pm 0.2204$ & +9.1\% \\
      Neural Demand (CF) & $-22.7475 \pm 0.1819$ & +4.3\% \\
      Neural Demand (habit, CF) & $-20.9201 \pm 0.1073$ & +11.9\% \\
      Neural Demand (habit, FE, CF) & $-21.8383 \pm 0.1277$ & +8.1\% \\
      PlaceboND & $-23.6393 \pm 0.0631$ & +0.5\% \\
      \bottomrule
    \end{tabular}
    \begin{tablenotes}\small
      \item Compensating variation via 100-step Riemann sum, $p_{Aspirin}\to(1+0.1)\,p_{Aspirin}$.
    \end{tablenotes}
  \end{threeparttable}
\end{table}

\subsubsection{Acetaminophen price increase}

Table~\ref{tab:dom_welfare_acetaminophen} reports CV losses from a 10\% acetaminophen price
increase. The habit-augmented models again imply smaller losses than the static neural benchmark
(roughly $+4$--$6\%$), but the dispersion across models is more moderate than for the ibuprofen shock.
This is consistent with acetaminophen's role as a close substitute with ibuprofen: models differ in how
much substitution and cross-price reallocation they permit, which affects the compensating payment.
As above, the placebo model remains close to the static benchmark, supporting the interpretation that
differences between static and habit welfare conclusions reflect genuine state dependence rather than a
generic ``more covariates'' effect.

\begin{table}[htbp]
  \centering
  \caption{Welfare Loss from 10\% Acetaminophen Price Increase --- Dominick\'s Analgesics (5 run(s); mean $\pm$ std)}
  \label{tab:dom_welfare_acetaminophen}
  \begin{threeparttable}
    \begin{tabular}{lcr}
      \toprule
      \textbf{Model} & \textbf{CV Loss (\$)} & \textbf{vs Neural Demand (static)} \\
      \midrule
      LA-AIDS & $-53.2076 $ & -2.8\% \\
      BLP (IV) & $-47.9725 $ & +7.3\% \\
      QUAIDS & $-52.9144 $ & -2.2\% \\
      Series Est. & $-51.2835 $ & +1.0\% \\
      Linear Demand (Shared) & $-46.4907 $ & +10.2\% \\
      Linear Demand (GoodSpec) & $-46.5685 $ & +10.1\% \\
      Linear Demand (Orth) & $-56.5144 $ & -9.2\% \\
      Neural Demand (static) & $-51.7759 \pm 0.4461$ &  \\
      Neural Demand (window) & $-56.8681 \pm 0.1239$ & -9.8\% \\
      Neural Demand (habit) & $-49.6957 \pm 0.1801$ & +4.0\% \\
      Neural Demand (FE) & $-49.7675 \pm 0.8430$ & +3.9\% \\
      Neural Demand (habit, FE) & $-48.9284 \pm 0.1360$ & +5.5\% \\
      Neural Demand (CF) & $-50.8257 \pm 0.4266$ & +1.8\% \\
      Neural Demand (habit, CF) & $-49.2605 \pm 0.2414$ & +4.9\% \\
      Neural Demand (habit, FE, CF) & $-48.5490 \pm 0.2209$ & +6.2\% \\
      PlaceboND & $-52.6722 \pm 0.4248$ & -1.7\% \\
      \bottomrule
    \end{tabular}
    \begin{tablenotes}\small
      \item Compensating variation via 100-step Riemann sum, $p_{Acetaminophen}\to(1+0.1)\,p_{Acetaminophen}$.
    \end{tablenotes}
  \end{threeparttable}
\end{table}

\end{document}